\documentclass[12pt,a4paper]{article}
\pdfoutput=1
\usepackage{a4wide,latexsym,graphicx,epsfig,psfrag,float}
\usepackage[english]{babel}
\usepackage{hyperref}
\usepackage{amsmath}
\usepackage{amssymb}
\usepackage{dsfont}
\usepackage{mathrsfs}
\usepackage{slashed}
\usepackage{longtable}
\usepackage{enumerate}
\usepackage{cite}

\allowdisplaybreaks

\begin{document}
\parskip=3pt plus 1pt

\begin{titlepage}
\begin{flushright}
{
IFIC/14-74
}
\end{flushright}
\vskip 1cm

\setcounter{footnote}{0}
\renewcommand{\thefootnote}{\fnsymbol{footnote}}
\vspace*{2cm} 
\begin{center}
{\LARGE \bf Instanton-mediated baryon number violation}
\\ [13pt]{\LARGE \bf in non-universal gauge extended models}
\vspace{1.5cm} \\
{\sc  J.~Fuentes-Mart\'{\i}n}\footnote{Email:~Javier.Fuentes@ific.uv.es}
{\sc , J.~Portol\'es}\footnote{Email:~Jorge.Portoles@ific.uv.es} and
{\sc P.~Ruiz-Femen\'ia}\footnote{Email:~Pedro.Ruiz@ific.uv.es}
\vspace{1.2cm} \\
Instituto de F\'{\i}sica Corpuscular, CSIC - Universitat de Val\`encia, \\
Apt. Correus 22085, E-46071 Val\`encia, Spain
\end{center}

\setcounter{footnote}{0}
\renewcommand{\thefootnote}{\arabic{footnote}}
\vspace*{1.5cm}

\begin{abstract}
Instanton solutions of non-abelian Yang-Mills theories generate an effective action that may induce lepton and baryon number
violations, namely $\Delta B = \Delta L = n_f$, being $n_f$ the number of families coupled to the gauge group. In this 
article we study instanton mediated processes in a $SU(2)_{\ell} \otimes SU(2)_h \otimes U(1)$ extension of the Standard Model that breaks universality by singularizing the third  family. In the construction of the instanton Green functions we account systematically for the inter-family mixing.
This allows us to use the experimental bounds on proton decay in order to constrain the gauge coupling of $SU(2)_h$. 
Tau lepton non-leptonic and radiative decays with $\Delta B = \Delta L = 1$ are also analysed. 
\end{abstract}
\vfill
PACS~: 11.30.Fs, 11.30.Hv, 12.60.Cn, 13.35.Dx \\
Keywords~: Baryon Number Violation, Lepton Number Violation, Proton Decay, Tau Decays

\end{titlepage}

\section{Introduction}
\label{sec:intro}
The undisputed success of the LHC and its dedicated experiments in the first period of runs at $\sqrt{E} \sim 7 \, \mbox{TeV}$ has provided plenty of data whose analyses are reinforcing the solidity of the Standard Model (SM). The discovery of the Higgs boson, the apparent lack of supersymmetric particles, together with the 
high precision achieved in many observables, still agreeing with SM predictions, confirm that deviations from the SM at present energies (not considering the issue of the neutrino masses) seem tiny.
However our present understanding of the structure of Nature in the realm of particle physics lets us expect that as we go back in time, with the Universe getting hotter and symmetries being restored, new symmetries and new spectra, which include the SM features, should appear. This New Physics could be around the corner, at the reach of the LHC or the super-B factories.
\par
A remarkable contradiction happens with the baryon number symmetry ($B$) of the SM Lagrangian and the apparent huge baryon asymmetry of our Universe, i.e. the preponderance of matter over antimatter. 
Some models of particle physics extending the present framework (Beyond the Standard Model) try to include the violation of baryon number symmetry or, for that matter, of other unprotected global symmetries of the SM like lepton number ($L$), but keeping $B-L$ as a symmetry of the theory. This has a 
resemblance with the peculiar status of these symmetries in the SM to which we now turn to. 
\par
The Standard Model Lagrangian has a $U(1)_B \otimes U(1)_e \otimes U(1)_\mu \otimes U(1)_\tau$ global symmetry.
On one side the fact that there is neutrino flavour mixing already points out that the global symmetry applied to lepton flavours
is no longer appropriate and it opens the interesting hunt for theorizing and observing charged lepton flavour violation. 
The $U(1)_{B+L}$ subgroup, on the other side, is anomalous
i.e. for the associated currents, $\partial_{\mu} {\cal J}_{B}^{\mu} = \partial_{\mu} {\cal J}_{L}^{\mu} = {\cal O}(\hbar)$.
Hence $B$ and $L$ are symmetries of the classical SM Lagrangian but quantum effects provide deviations, though conserving $B-L$.
The breaking of those global symmetries is of non-perturbative nature in the SM and has to do with the interaction
 of fermions generated through tunnelling between different vacua produced by instanton solutions of the Yang-Mills theory \cite{Polyakov:1975rs,Belavin:1975fg,'tHooft:1976up,'tHooft:1976fv}. 
 Each instanton transition between $SU(2)_L$ gauge vacua gives $\Delta B = \Delta L = n_f$, with $n_f$ the number of families or
generations that transform non trivially under the gauge group. However these transitions are enormously
suppressed at zero temperature by a factor ${\cal O}\left( \exp \left[ - 8 \pi^2/g^2 \right] \right)$, being 
$g \simeq \sqrt{4 \pi \alpha_{\mbox{\tiny{em}}}}/\sin \theta_{\mbox{\tiny{W}}} \sim 0.6$ the $SU(2)_L$ coupling. Effectively $B$ and $L$ symmetries turn out to be almost exact in the SM. 
\par
Hence it is clear that the observation of processes that deviate from these global symmetries would be an excellent
opening into New Physics. Here, and after the later results by LHCb \cite{Aaij:2013fia}, we tackle the study of 
decays of the tau lepton with $\Delta B = \Delta L = 1$. The tau lepton, the only lepton that is able to decay into hadrons, provides an excellent benchmark for key particle physics issues like hadronization of the QCD currents, Higgs physics,
tests of universality of the gauge couplings, determination of $\alpha_{\mbox{\tiny S}}$, lepton flavour violation, etc. \cite{Pich:2013lsa}. Both LHCb and future Super-B factories like Belle~II are, in fact, tau factories too and they have ambitious work programmes on tau physics. Motivated by present and future data on bounds for tau decays that violate $B$ and $L$ we study here processes like $\tau \rightarrow p \gamma$, $\tau \rightarrow p \mu^+ \mu^-$, $\tau \rightarrow p \pi^0$, etc. in a gauge extended version of the SM. These processes are severely constrained by proton decay as argued in \cite{Marciano:1994bg}, a correlation that is further analysed in this work. 
\par
From the discussion above one could conclude that the rate for these instanton-generated $B+L$ violating processes
could be much larger if the Yang-Mills coupling $g$ was not so small. Hence a possible gauge extension of the SM involving 
an additional $SU(2)$ group, with a larger coupling, could provide an appropriate framework to study those processes. Indeed these models constitute one of the simplest extensions of the electroweak gauge symmetry and are generically denoted as $G(221)$ models\cite{Li:1981nk,Chivukula:2003wj,Hsieh:2010zr}. We choose one of these as our playing ground. Moreover we will be interested in the analysis of the phenomenology of processes with $\Delta B = \Delta L = 1$. This constraint imposes a specific structure on the settings of the extended model. As we pointed out before, instanton-generated processes in the SM provide transitions with
$\Delta B = \Delta L = n_f$, i.e. equal to the number of families coupled to the Yang-Mills gauge group. In the SM due to universality of the couplings of matter to the gauge bosons, $n_f = 3$. Accordingly we need to extend the SM by breaking such universality: we need a model that couples the third family to a $SU(2)$ group, while the other two families are singlets.
This model has already been studied \cite{Muller:1996dj,Malkawi:1996fs,Chiang:2009kb,Hsieh:2010zr} and we collect its essentials in Section~\ref{sec:model}. In fact
the analysis of $\Delta B = \Delta L =1$ processes in this framework has already been considered in Ref.~\cite{Morrissey:2005uza}, where the authors used the instantonic effective interactions to constrain the gauge coupling. However, and as far as we know, neither in this reference nor in those that consider the instanton-generated effective action in the SM, a systematic study of the inter-family mixing has been considered. In this article we perform this task. We will show that, specially for baryon and lepton number violating processes involving the first and second families, the inclusion of the inter-family mixing is crucial. For instance, in the approach of Ref.~\cite{Morrissey:2005uza} the process $p \rightarrow K^+ \overline{\nu}_{\tau}$ was computed inserting the instanton-generated effective operator into a two-loop diagram, which gives rise to a cut-off dependence from the loop integrals. Conversely, the same process would appear in our framework as a tree-level insertion of the instanton-generated operator.
\par
In Section~\ref{sec:B+Llag} we recall (with the help of Appendices~\ref{ap:GreenFunc} and~\ref{ap:mixing}) the construction of the instanton-generated effective action  within our theoretical framework, with much emphasis in the determination of the fermionic zero modes when peculiarities arisen from flavour mixing are taken into account. Appendices~\ref{ap:projectors} and \ref{ap:gaugeInt} collect several technicalities regarding the determination of the zero modes. The constraints imposed by proton decay together with the results of $\Delta L=\Delta B=1$ tau decays are pointed out in Section~\ref{sec:protondec} while the construction of the $\Delta L=\Delta B=1$ effective chiral Lagrangian necessary to compute these observables is detailed in Appendix~\ref{ap:taudec}. We end with Conclusions in Section~\ref{sec:conclusions}.

\section{The \texorpdfstring{$\boldsymbol{SU(2)_{l}\otimes SU(2)_h \otimes U(1)_Y}$}{} model}
\label{sec:model}
A detailed study of models of extended electroweak gauge symmetry incorporating replicated $SU(2)$ and $U(1)$ gauge groups has been done in Ref.~\cite{Chivukula:2003wj}. These are among the simplest extensions of the SM. Particular interest has been arisen by the $SU(2) \otimes SU(2) \otimes U(1)$ models known as $G(221)$. Their characteristic feature is the inclusion of three new heavy gauge bosons, $W'^{\pm}$ and $Z'$, and the phenomenology generated by the new dynamics depends on both the specific symmetry breaking schemes as well as the charge assignments to fermions and scalars.
\par
We will consider a theory with an electroweak group ${\cal G} \equiv SU(2)_{l} \otimes SU(2)_h \otimes U(1)_Y$ that embeds the Standard Model and gives a general good description at the scale of a few TeV \cite{Chiang:2009kb}. As commented above we are interested in the study of instanton-generated $\Delta B = \Delta L = 1$ processes and this requires to singularize one of the families.
The fermion content of the model is the same as in the SM with the $SU(2)_L$ doublets of the first two families, both quark ($Q$) and leptons ($L$) being doublets under the $SU(2)_{l}$ group and singlets under $SU(2)_h$, while the third family reverses this assignment. $SU(2)_L$ singlets ($u$,$d$,$e$) remain singlets under the new dynamics. The scalar sector is slightly more complicated. There are two Higgs doublets, $\Phi_{l}$ for $SU(2)_{l}$ and $\Phi_h$ for $SU(2)_h$, which give masses to the fermions of the first and second families and to the third one, respectively. In addition, a self-dual bi-doublet scalar $\Phi_b$, such that $\Phi_b=\tau_2 \Phi_b^*\tau_2$ being $\tau_2$ the Pauli matrix, is introduced in order to recover the SM gauge symmetry via spontaneous symmetry breaking. 
In summary we have for the fermions:
\begin{align} \label{eq:frepr}
\begin{aligned}
Q_{i}&:\;\left(2,1\right)\left(1/3\right), &\qquad Q_{3}&:\;\left(1,2\right)\left(1/3\right),  \\
L_{i}&:\;\left(2,1\right)\left(-1\right), &\qquad L_{3}&:\;\left(1,2\right)\left(-1\right),  \\
u_{j}&:\;\left(1,1\right)\left(4/3\right), &\qquad d_{j}&:\;\left(1,1\right)\left(-2/3\right),  \\
e_{j}&:\;\left(1,1\right)\left(-2\right), &\qquad  
\end{aligned}
\end{align}
and
\begin{align} \label{eq:hrepr}
\begin{aligned}
\Phi_l&:\;\left(2,1\right)\left(1\right), &\qquad \Phi_h&:\;\left(1,2\right)\left(1\right),  \\
\Phi_b&:\;\left(2,2\right)(0), & 
\end{aligned}
\end{align}
for the scalars,
where the two numbers in the first parenthesis indicates the representation under $SU(2)_l \otimes SU(2)_h$ while the third number stands for the $U(1)$ hypercharge and $i=1,2$ and $j=1,2,3$ are family indices. The components of the bi-doublet are defined as:
\begin{align}
\Phi_b=\frac{1}{\sqrt{2}}\begin{pmatrix}
                         \Phi_b^0 & \Phi_b^+\\
                         -\Phi_b^- & \overline{\Phi}_b^0
                         \end{pmatrix}\,.
\end{align}
The symmetry breaking of $SU(2)_l \otimes SU(2)_h \to SU(2)_L$ is realized by the vacuum expectation value $\Phi_b \rightarrow \langle \Phi_b \rangle = u/2 \; \mathbb{I}$. The expected energy scale at which this happens is around $E \sim u \sim$ a few TeVs. This corresponds to the Non-Universal (NU) $G(221)$ model with scheme II of symmetry breaking pattern introduced in Ref.~\cite{Hsieh:2010zr}. In this scheme and for natural values of the parameters, the bi-doublet gets a mass of order $u$ and almost decouples from the other doublets, leaving a two-Higgs-doublet model at the electroweak scale. Finally, the electroweak SM symmetry is broken down to electromagnetism at the energy scale $E \sim \langle \Phi^0 \rangle = v/\sqrt{2} \simeq 0.174 \, \mbox{TeV}$, where $\Phi$ is the doublet that takes vacuum expectation value in the Higgs basis which emerges as a linear combination of the two Higgs doublets $\Phi_l$ and $\Phi_h$.
\par
The Lagrangian of the electroweak model is given by \cite{Chiang:2009kb}:
\begin{align} \label{eq:lagrange}
\begin{aligned}
{\cal L}  =&  - \frac{1}{4} \sum_{a=1}^3 W_{l \, \mu \nu}^a W_{l}^{a \, \mu \nu} \, - \, \frac{1}{4} \sum_{a=1}^3 
W_{h \, \mu \nu }^a W_h^{a \, \mu \nu} \, - \, \frac{1}{4} B_{\mu \nu} B^{\mu \nu} \, \\
& + \, \sum_{k=1}^3  i \, \psi_{L\,k}^{\dagger}  \, \overline{\sigma}^{\mu} \,  D_{\mu} \,  \psi_{L\,k} \, 
+ \, \sum_{j=1}^3  i \, \psi_{R\,j}^{\dagger} \,  \sigma^{\mu} \,  D_{\mu} \,  \psi_{R\,j} \, + \, {\cal L}_{\mbox{\tiny{Y}}}\\
& + \, \sum_{r=h,l} \, 
\left( D_{\mu} \Phi_r \right)^{\dagger} \left( D^{\mu} \Phi_r \right) \, + \, \mbox{Tr} \left[ \left(D_{\mu} \Phi_b \right)^{\dagger} \left( D^{\mu} \Phi_b \right) \right] \,  \, - \, V\left[\Phi_s \right] ,
\end{aligned}
\end{align}
where $k$ and $j$ are family indices and $\psi_{L\,k}$ denotes left-handed $SU(2)$ fermion doublets both quarks ($Q_k$) and leptons ($L_k$), while $\psi_{R\,k}$ stands for right-handed $SU(2)$ singlet quarks ($u_j, d_j$) and charged leptons ($e_j$), all of them in the Weyl 
representation. Here $W_{l,h \, \mu \nu}^a$ and $B_{\mu \nu}$ are the
gauge bosons field strength tensors, $\sigma^{\mu} = \left(1, \vec \sigma \right)$ and $\overline{\sigma}^{\mu} = \left( 1, - \vec \sigma \right)$
($\sigma^i$ are the Pauli matrices acting on the Weyl spinor space) and $V\left[\Phi_s \right]$ is the
higgses potential with $\Phi_s$ generically denoting all the scalars of the model, i.e. $s=l,\,h,\,b$.
The gauge dynamics of the model is encoded in the covariant derivatives acting on fermion fields and higgses:
\begin{align} \label{eq:dcov}
\begin{aligned}
D^{\mu} \, \psi_{L \, k} \, =& \, \left( \partial^{\mu} - i g_h  W_h^{\mu} \, \delta_{k 3} -i g_{l}  W_{l}^{\mu} \, \delta_{k \{1,2\}} - i \frac{g'}{2} Y B^{\mu} \right) \, \psi_{L \, k} , \\
D^{\mu} \, \psi_{R \, j} \, =& \, \left( \partial^{\mu} - i g' Q B^{\mu} \right) \, \psi_{R \, j}, \\
D^{\mu} \, \Phi_r \, =& \, \left( \partial^{\mu} - i g_h  W_h^{\mu} \, \delta_{r h} -i g_{l}  W_{l}^{\mu} \, \delta_{r l} - i \frac{g'}{2}  Y  B^{\mu} \right) \, \Phi_r , \; \; \; \; r=l,h, \\[0.3cm]
D^{\mu} \, \Phi_b \, =& \,  \partial^{\mu} \, \Phi_b \, + \,  i g_h \, W_h^{\mu} \, \Phi_b - \, i g_{l} \,  W_{l}^{\mu} \, \Phi_b ,
\end{aligned}
\end{align}
indices $k$, $j$ are as in Eq.~(\ref{eq:lagrange}) and $W_r^{\mu} = \tau^a \, W_r^{\mu a}/2$ for $r=l,h$ ($\tau^a$ are the Pauli matrices on the $SU(2)$ space). The hypercharge $Y$ is the $U(1)_Y$ SM quantum number that satisfies $Q = \tau^3/2 + Y/2$ and $Y=1$ for both $\Phi_h$ and $\Phi_l$ Higgs fields while the bi-doublet has zero hypercharge. The other fields have the same hypercharge as in the SM. The Yukawa Lagrangian in Eq.~(\ref{eq:lagrange}) is 
given by:
\begin{align} \label{eq:yukawa}
\begin{aligned}
- {\cal L}_{\mbox{\tiny{Y}}} = &\;  Y_{ij}^l \, u^{\dagger}_{i} \, \tilde{\Phi}_l^{\dagger} \, Q_{j} \, + \, 
Y_{i3}^h \,  u^{\dagger}_{i} \, \tilde{\Phi}^{\dagger}_h \, Q_{3} \, + \, 
X_{ij}^l \, d^{\dagger}_{i} \, \Phi_l^{\dagger} \, Q_{j} \, + \, 
X_{i3}^h \, d^{\dagger}_{i} \, \Phi^{\dagger}_h \, Q_{3} \, \\
& + \, Z_{ij}^l \, e^{\dagger}_{i} \, \Phi_l^{\dagger} \, L_{j} \, +
\, Z_{i3}^h \, e^{\dagger}_{i} \, \Phi_h^{\dagger} \, L_{3} \, + \, h.c. ,
\end{aligned}
\end{align}
with $\tilde{\Phi} = \varepsilon \Phi^{*} \equiv i \tau^2 \Phi^{*}$. Notice that here $i = 1,2,3$ and $j=1,2$ indicate the families. As we will assume that CP is a symmetry of our theory, the phases of the fermionic fields can be chosen so that the Yukawa couplings $X^{l,h}$, $Y^{l,h}$ and $Z^{l,h}$ are real.
\par 
After the last spontaneous symmetry breaking the charged gauge bosons $W_{l}$ and $W_h$ are not the physical states and a diagonalization procedure is necessary. Finally the mass eigenstates turn out to be $W$ and $W'$ with masses:
\begin{align} \label{eq:masw}
M_{\mbox{\tiny W}}^2 \, = \, \frac{1}{4} \, g^2 \, v^2 + {\cal O}(x^2) \; & , & \; 
M_{\mbox{\tiny W}'}^2 \, = \, \frac{1}{2} \, \left( g_h^2 +  g_{l}^2 \right) \, u^2 + {\cal O}(x^2) , 
\end{align}
where $x=v/u$ and $g = g_h g_{l}/\sqrt{g_h^2 + g_{l}^2}$ is the $SU(2)_L$ coupling. From Eq.~(\ref{eq:masw}) it can be concluded that there is a light gauge boson eigenstate, corresponding to the SM one, and a heavier gauge boson whose mass is proportional to the vacuum expectation value of the bi-doublet scalar after the first spontaneous symmetry breaking, $u$, since $g_h^2+g_l^2>g^2$. An analogous setting happens for the neutral $Z$ and $Z'$ bosons. Notice that when either $g_h$ or $g_l$ becomes large, the other one approaches $g$ from above, and hence $g_l, g_h > g$. 
\par
The hunt for heavy $W'$ and $Z'$-like gauge bosons at the LHC could provide key information on extensions of the SM if they are at reach of the collider. Its discovery potential in $G(221)$ models has been considered in Refs.~\cite{Cao:2012ng,Kim:2014afa,Edelhauser:2014yra}. Moreover the study of the  measurement of correlated observables aiming to distinguish between different versions of these models has also been carried out \cite{Jezo:2012rm}. It can also be seen that the NU model is the only anomaly-free $G(221)$ with symmetry breaking pattern $SU(2)_l \otimes SU(2)_h \longrightarrow SU(2)_L$ that gives a rather high lower bound for the new gauge bosons, namely $M_{W'} \simeq M_{Z'} \gtrsim 3.6 \, \mbox{TeV}$ for the most reasonable range of parameters \cite{Hsieh:2010zr}. Other $G(221)$ models accommodate lighter gauge bosons with lower bounds around $M_{W'} \sim 0.3-0.6 \, \mbox{TeV}$ and $M_{Z'} \sim 1.7 \, \mbox{TeV}$, the latter being mostly enforced by flavour changing neutral currents constraints. 
\par
Strongly correlated with the masses of the new gauge bosons are the $g_l$ and $g_h$ couplings. From a best fit to electroweak precision observables, Ref.~\cite{Chiang:2009kb} concluded a value of $M_{W'} \simeq 2.8 \, \mbox{TeV}$ for $g_l \simeq 1$ and 
$g_h \simeq 0.8$, with no errors attached. The analysis of Ref.~\cite{Hsieh:2010zr} points out that the ratio $g_h/g_l$ is basically unconstrained. Finally \cite{Cao:2012ng,Kim:2011qk} indicate that for $M_{W'} \gtrsim 2.5 \, \mbox{TeV}$ one can 
accommodate $g_h/g_l \gtrsim 1$.  

\section{Instanton-mediated \texorpdfstring{$\boldsymbol{B+L}$}{B+L} violating Lagrangian}
\label{sec:B+Llag}
In four-dimensional non-abelian Yang Mills theories, there exist non-trivial solutions to the Euler-Lagrange equations which locally minimize the Euclidean action \cite{Polyakov:1975rs,Belavin:1975fg}. In general these solutions, called \textit{instantons}, are stable structures localized in space and (imaginary) time, and are defined as solutions of the classical field equations in Euclidean space that have a finite action. As a result, for $t \rightarrow \pm \infty$ the instanton must approach classical vacuum solutions of the theory. Then it differs from the vacuum solution only for a certain interval of time.
\par
In quantum field theory formulated in Minkowski space-time the instantonic solutions are defined as an analytic continuation from those of the Euclidean theory; these configurations generate non-trivial Green functions. In Minkowski space-time, instantons provide tunnelling transitions between different topologically inequivalent vacua of the system described by the Lagrangian. These transitions introduce a peculiar dynamical interaction when fermions are coupled to the gauge fields that may give rise to a violation of $B+L$. Performing a semi-classical expansion around these configurations should provide a good approximation to the solutions of the physical system. 
 The procedure that yields the lowest order instanton-generated Green function for a $SU(2)$ gauge theory with a general matter content has been sketched in Appendix~\ref{ap:GreenFunc} where we follow the discussion of Refs.~\cite{'tHooft:1976fv,'tHooft:1976up}. The main result is given by Eq.~(\ref{eq:GF2}):
\begin{align}\label{eq:iGF2}
\widetilde{G}\left(p_1,\dots, p_{N_f}\right)= \left(2\pi\right)^4\delta^4\left(\sum_{i=1}^{N_f} p_i\right) \, \int dU\int d\rho\,e^{-S_{\mbox{\tiny E}}\left[ A^I,\, \Phi^I_s \right]}
F\left(\rho;\mu\right)\prod_{i=1}^{N_f} \widetilde{\psi}_{0,i}\left(p_i\right).
\end{align}
Here $N_f$ is the number of fermion doublets coupled to the $SU(2)$ group,
$F(\rho;\mu)$ is given by Eq.~(\ref{eq:frhomu}) and $S_{\mbox{\tiny E}}\left[ A^I, \Phi^I_s \right]$ by Eq.~(\ref{eq:SCL}) while $\rho$ and $U$ are the instanton radius and gauge orientation, respectively. Finally $\widetilde{\psi}_{0,i}\left(p_i\right)$ are the Fourier transform of the zero modes of the fermion operator in the instanton background. Their computation is the subject of Subsection~\ref{subsec:B+Llag1}.
\par
We intend to study instanton transitions in the non-universal gauge extended model presented in Section~\ref{sec:model}. We will restrict to computations of the $SU(2)_h$ instantons by setting the other gauge couplings to zero. Hence we consider instanton-generated $\Delta B = \Delta L = 1$ processes that involve only the third family before mixing. This calculation has been previously considered in the literature \cite{Morrissey:2005uza}, where the mixing between quark families was not included when constructing the instantonic effective interaction. In this article we provide a setting that takes into account systematically the inter-family mixing. As a result, we show that the flavour structure of the gauge currents is inherited by the instantonic transitions. 
 
\subsection{Fermion zero modes in the instanton background}
\label{subsec:B+Llag1}
We consider here the zero modes associated to the third family, i.e. the one that transforms non trivially under $SU(2)_h$, both 
leptons and quarks. In this work we assume that there is inter-family mixing between the quark families but we assume that neutrinos are massless and, accordingly, there is no lepton-family mixing. Nonetheless, the inclusion of lepton mixing is straightforward from the computation below. As far as we know, the mixing between families of quarks has not been considered previously in the framework of instanton dynamics when solving the fermion zero modes. We tackle here
this goal and, for that purpose, we will only detail the procedure for the quark fields. Lepton zero modes can be calculated analogously.
\par
The computation will be performed in the Euclidean space. Hence we proceed with the Lagrangian presented in Eq.~(\ref{eq:lagrange})
but now in the Euclidean. In this space the $SO(3,1)$ group will be substituted by $SO(4)$ where the two spinor representations are not related by complex conjugation. The relation between them ($\chi_{A,B}$) and those of $SO(3,1)$ ($\psi_{L,R}$) is generically given by:
\begin{align} \label{eq:eucl}
\psi_R \, \rightarrow \, \chi_A, \;\;\;\; \;  \; \;  \psi_L \, \rightarrow \, \chi_B ,\;\; \; \; \; \;\; \psi_R^\dagger \, \rightarrow \, \chi_B^\dagger, \;\;\; \; \; \; \;   \psi_L^\dagger \, \rightarrow \, \chi_A^\dagger .
\end{align}
Using the relation between the Minkowski and Euclidean actions, $i S_{\mbox{\tiny M}} = - S_{\mbox{\tiny E}}$, the Euclidean
Lagrangian reads:
\begin{align} \label{eq:lagrangee}
\begin{aligned}
{\cal L}_{\mbox{\tiny E}}  =& \;  \frac{1}{4} \sum_{a=1}^3 W_{l \, \mu \nu}^a W_{l \, \mu \nu}^{a}  \, + \, \frac{1}{4} \sum_{a=1}^3 
W_{h \, \mu \nu }^a W_{h \, \mu \nu}^{a} \, + \, \frac{1}{4} B_{\mu \nu} B_{\mu \nu} \,  + \, \sum_{r=h,\,l} \, 
\left( D_{\mu} \Phi_r \right)^{\dagger} \left( D_{\mu} \Phi_r \right) \\
&  + \, \mbox{Tr} \left[ \left(D_{\mu} \Phi_b \right)^{\dagger} \left( D_{\mu} \Phi_b \right) \right] \,  + \, V\left[\Phi_s \right]  \, + \, {\cal L}_Y^{\mbox{\tiny{E}}}\,  + \,   i \, {L}^{\dagger}_A  \, \hat{\overline{\sigma}}_{\mu} \,  D_{\mu} \,  L_B \, + \, i \, e^{\dagger}_B \, \hat{\sigma}_{\mu} \,  D_{\mu} \, e_A  \\
&  
\,  +  \,  i \, Q_{A}^{\dagger} \,  \hat{\overline{\sigma}}_{\mu} \,  D_{\mu} \,  Q_{B} \, 
+  \,  i \, u_{B}^{\dagger} \,  \hat{\sigma}_{\mu} \,  D_{\mu} \,  u_{A}  \, 
+  \,  i \, d_{B}^{\dagger} \,  \hat{\sigma}_{\mu} \,  D_{\mu} \,  d_{A} \, , 
\end{aligned}
\end{align}
where now $\hat{\sigma}_{\mu} = - \left( \vec \sigma, i \right)$ and $\hat{\overline{\sigma}}_{\mu} = \left( \vec \sigma, -i \right)$. In the fermionic kinetic terms a sum over the three families is understood and the covariant derivative $D_{\mu}$ 
has been defined in Eq.~(\ref{eq:dcov}). The Yukawa term, using the notation in Eq.~(\ref{eq:yukawa}) and the Euclidean relation between the spinors in Eq.~(\ref{eq:eucl}), is given by:
\begin{align} \label{eq:yuke}
\begin{aligned}
{\cal L}_Y^{\mbox{\tiny E}} =&   \,  Y_{ij}^l \left( u_{B \, i}^\dagger \, \tilde{\Phi}_l^\dagger \, Q_{B \, j} \, + \, Q_{A \, j}^\dagger \, \tilde{\Phi}_l \, u_{A \, i}\right) \, + \, 
Y_{i3}^h \,  \left( u_{B \, i}^\dagger \, \tilde{\Phi}^\dagger_h \, Q_{B \, 3}\, + \, Q_{A \, 3}^\dagger \, \tilde{\Phi}_h \, u_{A \, i} \right) \\
& \, + \, 
X_{ij}^l \, \left( d_{B \, i}^\dagger \, \Phi_l^\dagger \, Q_{B \, j}\, + \,  Q_{A \, j}^\dagger \, \Phi_l \, d_{A \, i}
 \right)  \, + \, 
X_{i3}^h \, \left( d_{B \, i}^\dagger \, \Phi^{\dagger}_h \, Q_{B \, 3}\, + \, Q_{A \, 3}^\dagger \, \Phi_h \, d_{A \, i} \right)   \\
&  \, + \, Z_{ij}^l \, \left( e_{B \, i}^\dagger \, \Phi_l^\dagger \, L_{B \, j} \, + \, L_{A \, j}^{\dagger} \, \Phi_l \, e_{A \, i} \right) \, +
\, Z_{i3}^h  \, \left(  e_{B \, i}^\dagger \, \Phi_h^\dagger \, L_{B \, 3} \, + \, L_{A \, 3}^\dagger \, \Phi_h \, e_{A \, i}   \right) ,
\end{aligned}
\end{align}
where, we remind, $i=1,2,3$ and $j=1,2$ are family indices. As explained above we intend to solve the fermion zero modes associated to the third family of fermions (both quarks and leptons). However the Yukawa interaction mixes the third family with the other two, a feature that after spontaneous symmetry breaking and mass diagonalization, gives a rich flavour physics structure. 
\par
\subsubsection{The Standard Model case}
\label{subsec:SMzeromod}
Before providing the solution of the fermion zero modes in the NU $G(221)$ model, and in order to ease their determination, we will explain the solution for the SM case, as many features of the computation are shared in both models. We will get the zero-mode quark fields in the background provided by the instanton solutions of the $SU(2)$ gauge bosons and the SM Higgs that have been collected in Appendix~\ref{ap:GreenFunc}. Here we do not consider the mixing between the $W^0$ and $B$ gauge bosons due to a non-zero $\theta_W$. This has no effect in the instanton-mediated Green functions we are computing because they do not contain gauge bosons as asymptotic states. For the computation of Green functions with gauge bosons in the external states, corrections introduced by a non-zero $\theta_W$ have been taken into account in Ref.~\cite{Gibbs:1995xt}. 
\par
Let us consider the quark doublet $q_j$, the quark singlets $u_j$ and $d_j$, in the mass-diagonal basis,\footnote{Notice that the fermion fields are now different from the ones previously introduced and are related to those by a flavour rotation (see Eqs.~\eqref{eq:mass-basis}).} and the SM Higgs doublet $\Phi$. The index $j$ indicates the family. The equations of motion of the SM for the fermion fields read:
\begin{align}\label{eq:EOM_SM}
\begin{aligned}
i \,\hat{\overline{\sigma}}_{\mu} \tilde{D}_\mu \, q_{B \, j} \, + \, \lambda_{u_j} \, \varepsilon \, \Phi^* \, u_{A \, j} \, + \,\lambda_{d_j} \, \Phi \, d_{A \, j}  = & \; 0, \\
- \, \lambda_{u_j} \, \Phi^T \, \varepsilon \, q_{B \, j} \, + \, i\hat{\sigma}_\mu  \partial_\mu  u_{A \, j}  = & \; 0, \\
 \lambda_{d_j}\, \Phi^\dagger \,  q_{B \, j} \, + \, i\hat{\sigma}_\mu  \partial_\mu  d_{A \, j}  = & \; 0,
\end{aligned}
\end{align}
with no summation in the flavour index, $j$, and where $\lambda_{p_j} = m_{p_j}/\langle \Phi^0 \rangle$ for $p=u,d$ and $m_{u_j}$ and $m_{d_j}$ are up- and down-type quark masses.
The covariant derivative, $\tilde{D}_\mu$, is defined as:
\begin{align}\label{eq:dcovesp}
\begin{aligned}
\tilde{D}_{\mu} & = \, \partial_{\mu} \, - \, i \, \frac{g}{2} \, W_\mu^a \, \mathscr{F}_{\mbox{\tiny SM}}^\dagger \, \tau^a \, \mathscr{F}_{\mbox{\tiny SM}} \, 
- \, i \, \frac{g'}{2} \, Y \, B_{\mu} \\
&=\partial_\mu \, - \, i \, \frac{g}{2} \begin{pmatrix}
                                        W_\mu^0 & \sqrt{2}\, W_\mu^+ \, V_{\mbox{\tiny CKM}}\\
                                        \sqrt{2 \, }W_\mu^- \, V_{\mbox{\tiny CKM}}^\dagger & -W_\mu^0
                                        \end{pmatrix} \, 
                                        - \, i \, \frac{g'}{2} \, Y \, B_{\mu},
\end{aligned}                                        
\end{align}
with $V_{\mbox{\tiny CKM}}$ the Cabibbo-Kobayashi-Maskawa matrix and $W_\mu^\pm=\frac{1}{\sqrt{2}}\left(W_\mu^1\mp iW_\mu^2\right)$ the $SU(2)$ electrically-charged gauge boson. The unitary matrix $\mathscr{F}_{\mbox{\tiny SM}}$ takes care of the inter-family mixing provided by $V_{\mbox{\tiny CKM}}$ and is given by:
\begin{align} \label{eq:mm}
\mathscr{F}_{\mbox{\tiny SM}}=\begin{pmatrix}
            1 & 0\\
            0 & V_{\mbox{\tiny CKM}}
            \end{pmatrix}.
\end{align}
We have to use for the $SU(2)$ gauge and Higgs fields the instanton classical solutions (see Eqs.~(\ref{eq:AI}) and (\ref{eq:PhiI}) in  Appendix~\ref{ap:GreenFunc}). As there are no instanton solutions for abelian Yang-Mills groups (in four-dimensional flat space-time), we have $B_{\mu}^{I}(x) = 0$ in Eq.~(\ref{eq:dcovesp}). 
\par
In order to solve Eqs.~(\ref{eq:EOM_SM}) we use the ansatz:
\begin{align} \label{eq:ansatze}
\begin{aligned}
q_{B \, j}\left(x\right) = & \; x_\mu \, \hat{\sigma}_\mu \, \xi_{B \, j}\left(y\right), \\
u_{A \, j} = & \; u_{A \, j}\left(y\right), \\
d_{A \, j} = & \; d_{A \, j}\left(y\right),
\end{aligned}
\end{align}
with $y = x_{\mu} x_{\mu} = x^2$. Eq.~(\ref{eq:EOM_SM}) becomes:
\begin{align}\label{eq:EOM_SM+ansatz}
\begin{aligned}
\mathscr{D}_{\mbox{\tiny U}} \, \xi_{B \, j}(y) \, + \, i \, \lambda_{u_j} \, \varepsilon \, \Phi^* \, u_{A \, j}(y) \, + \, i \, \lambda_{d_j} \, \Phi \, d_{A \, j}(y) = & \; 0, \\
- \lambda_{u_j} \, \Phi^T \, \varepsilon \, \xi_{B \, j}(y) \, + \, 2i \, \left( u_{A \, j}(y) \right)' = & \; 0, \\
\lambda_{d_j}\, \Phi^\dagger \, \xi_{B \, j}(y) \, + \, 2i \, \left( d_{A \, j}(y) \right)' = & \; 0,
\end{aligned}
\end{align}
where again there is no summation in the flavour index, $j$, and with the new derivative defined as:
\begin{align} \label{eq:dmuchi}
\mathscr{D}_{\mbox{\tiny U}} \, \xi_{B \, j}(y) \, = \, 4 \, \xi_{B \, j}(y) \, + \, 2 \, y \, \left(\xi_{B \, j}(y)\right)' \, + \, y \, \mathscr{W}(y) \, \mathscr{F}_{\mbox{\tiny SM}}^\dagger \, U \, \left(\vec{\sigma} \, \cdot \, \vec{\tau}\right) \, U^{\dagger} \, \mathscr{F}_{\mbox{\tiny SM}}  \, \xi_{B \, j}(y).
\end{align}
The prime on the fields denotes $d/dy$ and $\frac{1}{2}gW_\mu^a  \, \tau^a=\mathscr{W} x_\nu\eta^a_{\mu\nu} \, U \,  \tau^a \,   U^\dagger$, where $U$ parameterizes the instanton gauge orientation and $\eta^a_{\mu\nu}$ is a t' Hooft symbol that relates the $SO(4)$ generators to the $SU(2)$ generators~\cite{'tHooft:1976fv}. 
\par
The solution to Eqs.~(\ref{eq:EOM_SM+ansatz})  has to be worked out in both the short-distance, $x\ll\rho$, and the long-distance, $x\gg\rho$, regimes. This solution can be expressed as a perturbative expansion in the parameter $\rho\langle\Phi^0\rangle$. We will only be concerned with the leading order of this expansion. To proceed we
take $\mathscr{W}=\mathscr{A}^I$ and $\Phi=\Phi^I$ (see Eq.~(\ref{eq:Arara}) and Eq.~(\ref{eq:PhiI}) with $q_s=1/2$).
For the short-distance regime ($x\ll \rho$) the solution to Eqs.~(\ref{eq:EOM_SM+ansatz}), at lowest order in $\rho\langle\Phi^0\rangle$, is given by (here and until the end of this section we will assume that there is no summation in $j$ while the index $k$ is always implicitly summed):
\begin{align} \label{eq:xiud}
\begin{aligned}
\xi_{B \, j}\left(x\right)  = & \;\frac{\sqrt{2}}{\pi}\frac{\rho^{3/2}}{x\left(x^2+\rho^2\right)^{3/2}} \, \mathscr{F}_{\mbox{\tiny SM}}^\dagger \, U \, \zeta_{s_j}, \\
u_{A \, j}\left(x\right)  = & - \frac{i}{2\pi} \, m_{u_j} \,  \frac{\rho^{3/2}}{x^2+\rho^2} \, U \, \chi_{u_j}
, \\
d_{A \, j}\left(x\right)  = & -\frac{i}{2\pi}\, \left(V_{\mbox{\tiny CKM}}^\dagger\right)_{jk}\, m_{d_j} \, \frac{ \rho^{3/2}}{x^2+\rho^2} \, U \, \chi_{d_k},
\end{aligned}
\end{align}
where the spinors are $\chi_{u_j} = (0,1)^T$ and $\chi_{d_j} = (-1,0)^T$ and are orthogonal in flavour space while $\zeta_{s_j}$ is the singlet in the coupled spin-isospin space, which satisfies $\left(\vec{\sigma}\cdot\vec{\tau}\right)\zeta_{s_j}=-3 \,\zeta_{s_j}$ and $\zeta_{s_j}^\dagger\zeta_{s_j}=1$, namely $\zeta_{s_j} = (0,1,-1,0)^T/\sqrt{2}$. The triplet, $\zeta_{t_j}$, which satisfies $\left(\vec{\sigma}\cdot\vec{\tau}\right)\zeta_{t_j}= \,\zeta_{t_j}$, cannot solve Eqs.~\eqref{eq:EOM_SM+ansatz}.
\par
At long distances ($x\gg \rho$),  $y\mathscr{A}^I\to0$ and $\Phi^I\to\langle\Phi\rangle$ and the solutions to 
Eqs.~(\ref{eq:EOM_SM+ansatz}) have been worked out in Ref.~\cite{Espinosa:1989qn}. The first order in the perturbative expansion in $\rho\langle\Phi^0\rangle$ for the long-distance regime reads:
\begin{align} \label{eq:ldsol}
\begin{aligned}
u_{A \, j}\left(x\right) & = \frac{i}{2\pi} \, \rho^{3/2} \,  m_{u_j}^2\, \frac{K_1\left(m_{u_j} \,x\right)}{x}U \, \chi_{u_j} , \\
u_{B \, j}\left(x\right) & = -\frac{1}{2\pi} \, \rho^{3/2} \,  m_{u_j}^2\, \frac{K_2\left(m_{u_j} \, x\right)}{x^2} \, x_\mu \hat{\sigma}_\mu \,  U \, 
\chi_{u_j}  , \\
d_{A \, j}\left(x\right) & = \frac{i}{2\pi} \, \rho^{3/2} \,  m_{d_j}^2 \,   \left(V_{\mbox{\tiny CKM}}^\dagger\right)_{jk}\, \frac{K_1\left(m_{d_j} 
x\right)}{x}U \, \chi_{d_k} , \\
d_{B \, j}\left(x\right) & = -\frac{1}{2\pi} \, \rho^{3/2} \,  m_{d_j}^2 \, \left(V_{\mbox{\tiny CKM}}^\dagger\right)_{jk} \, \frac{K_2\left(m_{d_j} 
x\right)}{x^2} \, x_\mu \, \hat{\sigma}_\mu \, U \, \chi_{d_k} ,
\end{aligned}
\end{align}
where we have matched the long-distance solution with the short-distance solution in Eqs. \eqref{eq:xiud} in order to determine the global factors.
\par
In order to calculate the baryon number violating amplitudes that derive from the instanton-generated Green function we need the singular piece of the Fourier transform of the zero modes. This singularity is a pole in $p^2=-m^2$, being $m$ the mass of the particle, and only depends on the long-distance expansion of the zero mode \cite{Espinosa:1989qn}:
\begin{align} \label{eq:ftzm}
\begin{aligned}
u_{A \, j} \left(p\right) & = 2\pi i \, \rho^{3/2} \, \frac{m_{u_j}}{p^2+m_{u_j}^2} \, U \, \chi_{u_j}, \\
u_{B \, j}\left(p\right) & = -2\pi i \, \rho^{3/2} \, \frac{p_\mu \, \hat{\sigma}_\mu}{p^2+m_{u_j}^2} \,  U \, \chi_{u_j}, \\
d_{A \, j}\left(p\right) & = 2\pi i \, \rho^{3/2} \, \left(V_{\mbox{\tiny CKM}}^\dagger\right)_{jk} \, \frac{m_{d_j}}{p^2+m_{d_j}^2}  \, U \, \chi_{d_k}, \\
d_{B \, j}\left(p\right) & = -2\pi i \, \rho^{3/2} \, \left(V_{\mbox{\tiny CKM}}^\dagger\right)_{jk} \, \frac{p_\mu \, \hat{\sigma}_\mu}{p^2+m_{d_j}^2} \,  U \, \chi_{d_k}.
\end{aligned}
\end{align}
Reverting to Minkowski space and assembling the Weyl spinors into a Dirac spinor in the Weyl basis we get:
\begin{align}\label{eq:ftzmm}
\begin{aligned}
u_j(p)& = -2\pi i \, \rho^{3/2} \, \frac{\slashed{p}+m_{u_j}}{p^2-m_{u_j}^2} \, \left(\begin{matrix}
                                                                 0\\
                                                                 U \, \chi_{u_j}
                                                                 \end{matrix}\right), \\
d_j(p)& = -2\pi i \, \rho^{3/2} \,  \left(V_{\mbox{\tiny CKM}}^\dagger\right)_{jk} \, \frac{\slashed{p}+m_{d_j}}{p^2-m_{d_j}^2} \, \left(\begin{matrix}
                                                                 0\\
                                                                 U \, \chi_{d_k}
                                                                 \end{matrix}\right).
\end{aligned}                                                                 
\end{align}
By amputating the propagators and putting the particles on-shell we finally obtain:
\begin{align} \label{eq:ftamp}
\begin{aligned}
\left[u_j(p)\right]_{\mbox{\tiny Amp}}&=-2\pi i \, \rho^{3/2} \, \left(\begin{matrix}
                                                              0\\
                                                              U \, \chi_{u_j}
                                                              \end{matrix}\right) , \\
\left[d_j(p)\right]_{\mbox{\tiny Amp}}&=-2\pi i \, \rho^{3/2} \,  \left(V_{\mbox{\tiny CKM}}^\dagger\right)_{jk} \, \left(\begin{matrix}
                                                                 0\\
                                                                 U \, \chi_{d_k}
                                                                 \end{matrix}\right).
\end{aligned}                                                                 
\end{align}
Leptonic zero modes are obtained from the previous result just by changing $u\to\nu$, $d\to e$ and $V_{\mbox{\tiny CKM}}\to I$ (we consider neutrinos
to be massless). 

\subsubsection{The Non-Universal \texorpdfstring{$\boldsymbol{G(221)}$}{G(221)} model case}
\label{subsec:NUzeromod}
Once we have recalled the SM result for the $SU(2)$ fermion zero modes, let us proceed with the Non-Universal $G(221)$ model presented in Section~\ref{sec:model}. As we pointed out the key feature of this model, for our interests, is the fact that
the third family of the SM couples to the $SU(2)_h$ group while the other two SM families couple to $SU(2)_l$, i.e. there is a breaking of universality in the dynamics of the fermions. As we are looking for instanton-generated $\Delta B = \Delta L =1$ processes we will study the Green functions associated to the instanton solution given by the classical field configuration $W_h^{cl}=W_h^I$ and $W_l^{cl}=B^{cl}=0$, as any other classical minimization solution would contribute to processes with $\Delta B = \Delta L >1$ (see Appendix~\ref{ap:mixing} for more details). For this instanton background only the $SU(2)_h$ group becomes relevant. Therefore, we intend to determine the fermion zero modes associated to the $SU(2)_h$ group, the third family fermion zero modes. 
\par
In our case the (Euclidean) equations of motion in a constrained instanton background and in the mass-diagonal basis take a similar form as in the SM (see Eqs.~(\ref{eq:EOM_SM})). However, two differences arise:
\begin{itemize}
 \item As the $SU(2)_h$ gauge group only couples to one family, we have to include a projector operator in the interacting part of the covariant derivative $\tilde{D}_\mu$ in Eq.~(\ref{eq:dcovesp}) that corresponds to the gauge boson $W_h^\mu$. In this scenario, that part takes the following form:
\begin{align}\label{eq:dcovnonu}
\begin{aligned}
\tilde{D}^\mu &=\partial^\mu-\frac{1}{2}i \, g_h \, W_{h}^{\mu \,a} \, \mathscr{F}^\dagger\, \tau^a \, \mathscr{F}  \, + \, ... \\
&=\partial^\mu-\frac{1}{2}i \, g_h \begin{pmatrix}
                                W_{h}^{\mu \,0} \, P_u & \sqrt{2} W_{h}^{\mu \, +} \, P_u \, V_{\mbox{\tiny CKM}} \, P_d\\
                             \sqrt{2}W_{h}^{\mu \, -} \, P_d \, V_{\mbox{\tiny CKM}}^\dagger \, P_u & -W_{h}^{\mu \,0} \,  P_d
                                        \end{pmatrix} \, + \, ... ,
\end{aligned}                                        
\end{align}
where $P_{u,d}$ project up-type and down-type quarks onto the family space which couples to the gauge boson; the definition of these projector operators is given in Appendix~\ref{ap:projectors}. Consequently:
\begin{align}\label{eq:M_final}
 \mathscr{F}=\begin{pmatrix}
             P_u & 0\\
             0 & V_{\mbox{\tiny CKM}}P_d
             \end{pmatrix}.            
\end{align}
As there are now projector operators, the matrix $\mathscr{F}$ is non-invertible. Nonetheless, the analogue to Eqs.~\eqref{eq:EOM_SM+ansatz} for this model can still be solved by making use of the relations \eqref{eq:relation}  for the quark projectors. These relations allow us to rewrite Eq.~\eqref{eq:M_final} as:
\begin{align}
 \mathscr{F}=\begin{pmatrix}
             P_u & 0\\
             0 & P_u\,V_{\mbox{\tiny CKM}}
             \end{pmatrix}=P_u\,\mathscr{F}_{\mbox{\tiny SM}}.            
\end{align}

Notice that as $P_u$ has rank one there is only one zero mode for all fermion flavours (see Appendix \ref{ap:projectors} for more details), in contrast to what happened in the SM case where we had one solution for each flavour. This is expected because in this model only one family couples to the gauge group $SU(2)_h$.

\item The classical solution for the Higgs field in the instanton background of $SU(2)_h$ is given by (\ref{eq:PhiI}):
\begin{align}
\label{eq:inshig}
 \Phi^I \left(x\right) & =   c_\beta \, \Phi_l^I(x) \, + \, s_\beta \, \Phi_h^I(x)  = 
  \begin{cases}
    \left[ c_\beta \, \langle \Phi_l^0 \rangle \, + \, s_\beta \, \left(\frac{x^2}{x^2+\rho^2}\right)^{\frac{1}{2}}   \langle \Phi_h^0 \rangle \right] \, \overline{h}, &  x\ll \rho , \\[0.6cm]
 \; \left[ c_\beta \, \langle \Phi_l^0 \rangle  \, + \, s_\beta \, \langle \Phi_h^0 \rangle \, \right] \, \overline{h}\;= \frac{v}{\sqrt{2}}\,\overline{h} , & x\gg \rho ,
\\
 \end{cases}
\end{align}
as $\Phi_l$ is a singlet and $\Phi_h$ a doublet under the $SU(2)_h$ group. Here $\tan \beta = \langle \Phi_h^0 \rangle / 
\langle \Phi_l^0 \rangle$. The solution for $\Phi^I$ is now different and, as a consequence,  the $\rho$-dependence of the short-distance expression will be modified but the long-distance one will keep the same form. As we are only concerned about the pole part which is dominated by the long-distance expansion, this difference with respect the SM case does not change the fermionic zero modes.
\end{itemize}
\par 
With these differences taken into consideration, the amputated fermion zero modes for the $SU(2)_h$ instantons are given by:
\begin{align} \label{eq:zmodes}
\begin{aligned}
 \left[u(p)\right]_{\mbox{\tiny Amp}}&=-2\pi i \, \rho^{3/2} \, \left(\begin{matrix}
                                                              0\\
                                                              U \, \chi_u^{\mbox{\tiny P}}
                                                              \end{matrix}\right) , \\
 \left[d(p)\right]_{\mbox{\tiny Amp}}&=-2\pi i \, \rho^{3/2} \, \left(\begin{matrix}
                                                                 0\\
                                                                 U \, \chi_d^{\mbox{\tiny P}}
                                                                 \end{matrix}\right) ,
\end{aligned}                                                                 
\end{align}
where the normalized projected spinors now take the form (note the implicit sum in the flavour index $i$):
\begin{align}
\begin{aligned}
\chi_u^{\mbox{\tiny P}}&=\left(V_u\right)_{i3}\chi_{u_i},\\
\chi_d^{\mbox{\tiny P}}&= \left( V_d \right)_{i3} \, \chi_{d_i} \, = \, \left(V_{\mbox{\tiny CKM}}^\dagger\right)_{im}\left(V_u\right)_{m3}\chi_{d_i}.
\end{aligned}
\end{align}
As in the SM case, the leptonic zero modes are obtained from the previous result just by changing $u\to\nu$, $d\to e$ and $V_{\mbox{\tiny CKM}}\to I$.

\subsection{Instanton-induced effective operators}
Let us finally write down in this section the effective operators that reproduce the one-instanton amplitude that derives from the Green function in Eq.~(\ref{eq:GF2}). In order to identify the structure
and coefficient of the operators, we need to  work out the remaining integrations in 
both instanton size and instanton group orientation  (see Appendix~\ref{ap:GreenFunc}).  
Upon substituting the amputated fermion zero modes calculated in the previous section the one-instanton amplitude takes the general form:
\begin{align}\label{eq:amplitude1}
A=& \, C \, g^{-8} \, e^{-\frac{8\pi^2}{g^2\left(\mu\right)}} \, \left(2\pi i \right)^{N_f}
\int d\rho\,e^{-4\pi^2\rho^2\mathcal{V}^2} \, \rho^{\frac{3N_f}{2}-5} \, \left( \mu \rho \right)^{\beta_1} \, \int dU \, \prod_{f=1}^{N_f}
\overline{\omega_{f}} \, \, \left(\begin{matrix}
                                                                 0\\
                                                                 U \chi_{f}
                                                                 \end{matrix}\right),
\end{align}
where the flavour structure is encoded in the flavoured spinor $\chi_{f}$. Here $\omega_f(p)$, for $f=u,d,\nu,e$, indicates the external-state polarization spinor whose flavour indices were omitted for the sake of simplicity.  The constant $C$ is given in Eq.~(\ref{eq:cthooft}). 
The integral over instanton size can be trivially performed and gives:
\begin{align} \label{eq:igamma}
\int_0^{\infty} d\rho\,e^{-4\pi^2\rho^2\mathcal{V}^2}\rho^{\frac{3N_f}{2}-5+\beta_1}=\frac{1}{2}\left(\frac{1}{4\pi^2\mathcal{V}^2}\right)^{\frac{3N_f}{4}+\frac{\beta_1}{2}-2}\Gamma\left(\frac{3N_f}{4}+\frac{\beta_1}{2}-2\right).
\end{align}
\par
Let us proceed now to work out the amplitude $A$ in Eq.~(\ref{eq:amplitude1}) in the Non-Universal $G(221)$ model case. We have one quark and one lepton family (the ones associated to $SU(2)_h$) and we also have to take into account that quarks have three colours. Therefore the simplest operator that one can consider in the
amplitude is one with three quarks and one lepton, i.e. $N_f=4$. In addition the structure in Eq.~(\ref{eq:amplitude1}) indicates that all fermions are left-handed, as expected for a gauge group which only couples to left-handed fermions. Upon integration of the instanton gauge orientation (see Appendix~\ref{ap:gaugeInt}) we get that the only possible structures are $uude$ and $udd\nu$,
which conserve electric charge and violate $B+L$ in one unit while conserving $B-L$: 
\begin{align} \label{eq:operatore}
\begin{aligned}
\int dU\prod_{f=1}^{4}\overline{\omega_f} \,\left(\begin{matrix}
                                                                 0\\
                                                                 U \, \chi_f^{\mbox{\tiny P}}
                                                                 \end{matrix}\right) \, = & \; 
\frac{1}{6} \, \epsilon_{\alpha\beta\gamma} \,\overline{\omega_{u_i}^\alpha} \,\left(V_u\right)_{i3}\left(V_{\mbox{\tiny CKM}}^\dagger\right)_{jm}\left(V_u\right)_{m3} \left(\omega_{d_j}^\beta\right)^C  \\
& \times \left[\overline{\omega_{u_k}^\gamma}\left(V_u\right)_{k3}\left(V_\ell\right)_{l3}
\, \omega_{e_l}^C - \, \overline{\omega_{d_k}^\gamma} \left(V_{\mbox{\tiny CKM}}^\dagger\right)_{kn}\left(V_u\right)_{n3}\, \left(V_\ell\right)_{l3} \omega_{\nu_l}^C \right] \, ,
\end{aligned}
\end{align}
where $\alpha, \beta$ and $\gamma$ are colour indices and $i,j,\,\dots$ are family indices.
\par
Putting everything together and including also the contribution from anti-instanton transitions, we get the following effective Lagrangian for the one-instanton amplitude (writing the fields explicitly):
\begin{align} \label{eq:lbl1}
\begin{aligned}
\mathcal{L}_{B+L}= &\; \frac{C}{12 \, g_h^8}  \mu^{\beta_1}  e^{-\frac{8\pi^2}{g_h\left(\mu\right)^2}}  \left( 2 \pi \right)^{2-\beta_1}  
\left(\frac{1}{\mathcal{V}^2}\right)^{1+\frac{\beta_1}{2}} \Gamma\left(1+\frac{\beta_1}{2}\right) \epsilon_{\alpha \beta \gamma} \overline{\left(u_{L \, i}^{\alpha}\right)^C}\, \left(V_u\right)_{i3}\left(V_{\mbox{\tiny CKM}}^\dagger\right)_{jm}\left(V_u\right)_{m3}d_{L \, j}^{\beta}\\
&\times \left[\overline{\left(u_{L \, k}^{\gamma}\right)^C}\, \left(V_u\right)_{k3}\left(V_\ell\right)_{l3}\, e_{L \, l} -  \overline{\left(d_{L \, k}^{\gamma}\right)^C}\, \left(V_{\mbox{\tiny CKM}}^\dagger\right)_{kn}\left(V_u\right)_{n3}\, \left(V_\ell\right)_{l3} \, \nu_{L \, l}\right] \, +h.c.
\end{aligned}
\end{align}
Here $\beta_1$ and $C$ are given by Eqs.~(\ref{eq:onelbeta}) and (\ref{eq:cthooft}), respectively, with $N_f=4$ and $N_S=2$ ($N_S$ is the number of scalars
coupled to the gauge fields; in our case we have one doublet and one self-dual bi-doublet, the latter counting as one). Finally $\mathcal{V}^2 = \sum_s q_s \langle \Phi_s^0 \rangle^2 = 1/2 \left[ \langle \Phi_2^0 \rangle^2 + u^2/2 \right] \simeq u^2/4$.
Notice that the term $\mu^{\beta_1}e^{-\frac{8\pi^2}{g_h\left(\mu\right)^2}}$ is renormalization group invariant at one-loop order. The factor $g_h^{-8}$, however, is expected to be renormalized by higher-loop effects.
\par
In order to make contact with the notation used in the literature, 
we can rewrite the baryon number violating Lagrangian as:
\begin{align} \label{eq:lblcano}
\mathcal{L}_{B+L}=\left(C_{LL}^e\right)_{ijkl}\left(\mathcal{O}_{LL}^e\right)_{ijkl}+\left(C_{LL}^\nu\right)_{ijkl}\left(\mathcal{O}_{LL}^\nu\right)_{ijkl}+h.c.,
\end{align}
with
\begin{subequations} \label{eq:operatora}
\begin{align}\label{eq:operatorae}
\left(\mathcal{O}_{LL}^e\right)_{ijkl}&=\epsilon_{\alpha\beta\gamma}\overline{\left(u_{L \, i}^{\alpha}\right)^C} d_{L\, j}^{\beta}
\overline{\left(u_{L\, k}^{\gamma}\right)^C} \, e_{L\, l},  \\\label{eq:operatoranu}
\left(\mathcal{O}_{LL}^\nu\right)_{ijkl}&=\epsilon_{\alpha\beta\gamma}\overline{\left(u_{L\, i}^{\alpha}\right)^C} \, d_{L \, j}^{\beta} \, 
\overline{\left(d_{L \, k}^{\gamma}\right)^C} \, \nu_{L\, l}.
\end{align}
\end{subequations}
The Wilson coefficients are defined as:
\begin{subequations}\label{eq:wilsono}
\begin{align}\label{eq:wilsonoe}
\left(C_{LL}^e\right)_{ijkl}&=\frac{C}{12 \, g_h^{8}}\, \mu^{\beta_1} \, e^{-\frac{8\pi^2}{g_h\left(\mu\right)^2}}\, \left(2\pi\right)^{2- \beta_1} \, \left(\frac{1}{\mathcal{V}^2}\right)^{1+\frac{\beta_1}{2}}\, \Gamma\left(1+\frac{\beta_1}{2}\right) \, \left(V^e\right)_{ijkl},\\\label{eq:wilsononu}
\left(C_{LL}^\nu\right)_{ijkl}&=- \, \frac{C}{12 \, g_h^{8}} \, \mu^{\beta_1}e^{-\frac{8\pi^2}{g_h\left(\mu\right)^2}} \, \left({2\pi}\right)^{2-\beta_1} \, \left(\frac{1}{\mathcal{V}^2}\right)^{1+\frac{\beta_1}{2}} \, \Gamma\left(1+\frac{\beta_1}{2}\right) \, \left(V^\nu\right)_{ijkl},
\end{align}
\end{subequations}
where, using the definition of the projectors in Appendix~\ref{ap:projectors},
\begin{subequations}\label{eq:vpqrs}
\begin{align}
\label{eq:vpqrse}
\left(V^e\right)_{ijkl} &= \left(V_u\right)_{i3}\left(V_{\mbox{\tiny CKM}}^\dagger\right)_{jm}\left(V_u\right)_{m3} \left(V_u\right)_{k3}\left(V_\ell\right)_{l3},\\[0.2cm]
\label{eq:vpqrsnu}
\left(V^\nu\right)_{ijkl} &= \left(V_u\right)_{i3}\left(V_{\mbox{\tiny CKM}}^\dagger\right)_{jm}\left(V_u\right)_{m3}\left(V_{\mbox{\tiny CKM}}^\dagger\right)_{kn}\left(V_u\right)_{n3}\, \left(V_\ell\right)_{l3}. 
\end{align}
\end{subequations}

\section{Proton decay and baryon number violating tau decays}
\label{sec:protondec}

\begin{figure}[t]
\centering
\includegraphics[width=0.6\textwidth]{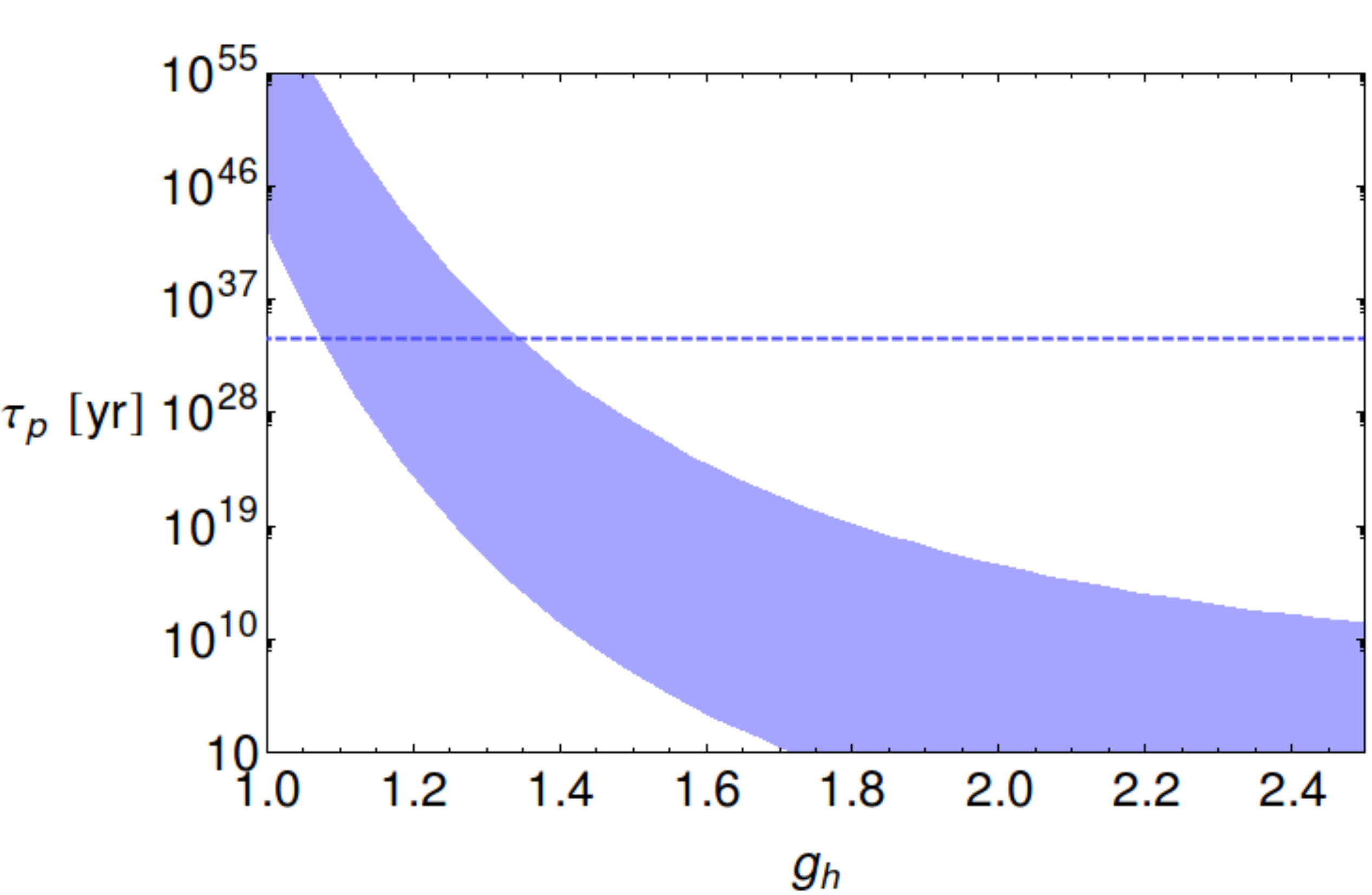}
\caption{Bounds on $g_h$ (the coupling of $SU(2)_h$) from $\tau_{p\to e^+\pi^0}$. 
The band represents the result in our model as a function of $g_h$ with $\mu=u=3$ TeV. 
The dashed line shows the lower bound on $p\to e^+\pi^0$ extracted
from the PDG~\cite{Beringer:1900zz}, $\tau_{p\to e^+\pi^0}> 8200\times 10^{30}$ years. 
The band arises from varying the flavour factor $V_{1111}^e$ in Eq.~(\ref{eq:vpqrse}) 
from $10^{-5}$ to $1$.}
\label{fig:protondec}
\end{figure}

In this section we provide constraints on the parameters of the baryon number violating
effective operators obtained in Section~\ref{sec:B+Llag}
using the current experimental bounds on proton decay. Baryon number violating decays of the tau lepton are also
considered.
\par
The decay widths of the $B+L$ violating proton decays into pseudoscalar mesons can be calculated using 
the formalism of chiral Lagrangians for baryon-meson strong interactions~\cite{Claudson:1981gh},
which is outlined in Appendix~\ref{ap:taudec}.
The analytic expressions for the
decays to one pseudoscalar meson and one lepton in the Born approximation
can be found in Ref.~\cite{Nath:2006ut}, parametrized in terms of the Wilson coefficients of the 
$\Delta B = \Delta L = 1$ dimension-6 operators. The decay mode with the largest partial mean life is 
$p\to e^+\pi^0$, with $\tau_{p\to e^+\pi^0}> 8200\times 10^{30}$ years~\cite{Beringer:1900zz}.
We can use this experimental
bound
and the result for the partial width~\cite{Nath:2006ut},
\begin{align}\label{eq:pdecay}
\begin{aligned}
\Gamma(p\to e^+\pi^0) = & \frac{(m_p^2-m_\pi^2)^2}{128\pi f_\pi^2\, m_p^3}\,
 \big| \beta (C_{LL}^e)_{1111} \big|^2 \, (1+D+F)^2 \\[2mm]
& \simeq (1.9\cdot 10^{-4}~\mbox{GeV}^5) \, \big|  (C_{LL}^e)_{1111} \big|^2
\,,
\end{aligned}
\end{align}
to constrain $| (C_{LL}^e)_{1111} |$:
\begin{align}
\big|  (C_{LL}^e)_{1111} \big| < 1.1\cdot 10^{-25}~\mbox{TeV}^{-2}
\,.
\label{eq:C1111bound}
\end{align}
Here $\beta$ is a hadronic parameter, defined in Eq.~(\ref{eq:strongpar}),
and the parameters $F$ and $D$ in Eq.~(\ref{eq:pdecay}) arise from the
baryon number conserving interactions defined in Eq.~(\ref{eq:L_B});
numerical estimates for the latter are given in Eqs.~(\ref{eq:strongparnum}) and~(\ref{eq:FDnum}),
respectively. The values for the rest of the parameters in Eq.~(\ref{eq:pdecay})
are specified in Appendix~\ref{ap:taudec}. The limit on $(C_{LL}^e)_{1111}$ above sets an upper bound on the value
of the coupling constant $g_h$ 
which enters in the coefficient, see Eq.~(\ref{eq:wilsonoe}).
This is shown in Figure~\ref{fig:protondec} where the band
accounts for the unknown value of the flavour factor $V_{1111}^e$ in Eq.~(\ref{eq:vpqrse}),
which satisfies $|V_{1111}^e|\leq 1$ due to the unitarity of the quark and lepton rotation 
matrices, but is unbounded from below. In Refs.~\cite{Lee:2004nz,Lee:2010zzq}, the author gave an estimate for the value of the matrices $V_u,\,V_d$ and $V_\ell$ from the phenomenological analysis of CKM unitarity violation and lepton flavour violation.\footnote{Notice the different convention in the definition of the matrices $V_u,\,V_d$ and $V_\ell$ between Refs.~\cite{Lee:2004nz,Lee:2010zzq} and the present work.} These analyses suggest that a value of $\left|V_{1111}^e\right|\simeq1$, for which $g_h<1.1$, can be easily accommodated with the current data. For the plot in Figure~\ref{fig:protondec} we have taken  
a conservative bound where $|V_{1111}^e|_{\rm min} = 10^{-5}$, which yields $g_h< 1.3$. 
The experimental bound for $p\to \mu^+\pi^0$ is slightly weaker than that of
$p\to e^+\pi^0$, and leads to a constraint on $(C_{LL}^e)_{1112}$ similar to that obtained
for $(C_{LL}^e)_{1111} $ in Eq.~(\ref{eq:C1111bound}).
\par
Since the proton cannot decay into $\tau$, setting direct limits on $(C_{LL}^e)_{1113}$
requires baryon number violating tau decays. Several $\Delta B=\Delta L=1$ tau decay rates
have been computed in Appendix~\ref{ap:taudec}. The corresponding branching fractions can be read off
the coefficient $a_3$ in Table~\ref{tab:rates}:
\begin{align}\label{eq:taudecaysnum}
{\cal B}(\tau^+\to p\,\pi^0) 
&\simeq (1.2\cdot 10^{-4}~\mbox{TeV}^4) \, \big| (C_{LL}^e)_{1113} \big|^2
\,,
\nonumber\\[2mm]
{\cal B}(\tau^+\to p\,\eta) &\simeq (8.6\cdot 10^{-5}~\mbox{TeV}^4) \, \big| (C_{LL}^e)_{1113} \big|^2
\,,
\nonumber\\[2mm]
{\cal B}(\tau^+\to \Lambda\,\pi^+) 
&\simeq (2.8\cdot 10^{-5}~\mbox{TeV}^4) \, \big| (C_{LL}^e)_{1113} \big|^2
\,,
\nonumber\\[2mm]
{\cal B}(\tau^+\to p\pi^0\pi^0) 
&\simeq (7.7\cdot 10^{-6}~\mbox{TeV}^4) \, \big| (C_{LL}^e)_{1113} \big|^2
\,,
\\[2mm]
{\cal B}(\tau^+\to p\pi^0\eta) &\simeq (1.2\cdot 10^{-6}~\mbox{TeV}^4) \, \big| (C_{LL}^e)_{1113} \big|^2
\,,
\nonumber\\[2mm]
{\cal B}(\tau^+\to p \gamma) &\simeq (2.3\cdot 10^{-7}~\mbox{TeV}^4) \, \big| (C_{LL}^e)_{1113} \big|^2
\,,
\nonumber\\[2mm]
{\cal B}(\tau^+\to p\,\mu^+\mu^-) &\simeq (7.9\cdot 10^{-10}~\mbox{TeV}^4) \, \big| (C_{LL}^e)_{1113} \big|^2
\,,\nonumber
\end{align}
where the last two processes involve electromagnetic radiation and are further suppressed by 
one a two powers of $\alpha_{em}$, respectively.
The experimental bounds on lepton and baryon number violating tau decays are however much weaker than those from proton
decay. The strongest bound comes from $\tau^-\to \bar{\Lambda} \pi^-$~\cite{Miyazaki:2005ng},
namely ${\cal B}(\tau^- \to \bar{\Lambda} \pi^-)< 1.4\times 10^{-7}$, 
equivalently $\tau_{\tau^-\to\bar{\Lambda} \pi^-} > 2.1 \cdot 10^{-6}$~s,
which is many orders of magnitude away from $\tau_{p\to e^+\pi^0}> 8200\times 10^{30}$ years.
Consequently, the best bound on $(C_{LL}^e)_{1113}$  obtained from hadronic tau decays is only 
$|(C_{LL}^e)_{1113}| < 0.7~\mbox{TeV}^{-2}$. Though future facilities like Belle II have a strong
physics programme on lepton flavour violation decays of the tau lepton, improvements on the experimental 
precision on baryon number violating tau decays are not foreseen at present. It is interesting to notice
that, at least in principle, low-energy hadron facilities could help to constraint those couplings through tau lepton production,
for instance, in pion-nucleon scattering, i.e. $\pi N \rightarrow \tau \pi$.
\par
An indirect way to have access to $(C_{LL}^e)_{1113}$ is through the 
$p\to \bar{\nu}_{\tau}\pi^+$ decay with a virtual $\tau$, as already suggested
in Ref.~\cite{Marciano:1994bg} (see also~\cite{Hou:2005iu}).
If no significant destructive interference ({\it i.e.} of orders of magnitude) 
between the direct $p\to \bar{\nu}_{\tau}\pi^+$  amplitude and the 
$p\to\tau^+\to \bar{\nu}_{\tau}\pi^+$ one is present, then we can argue that the decay rate given just by the 
latter 
must satisfy the experimental bound on $p\to \bar{\nu}_{\tau}\pi^+$ by itself. The computation of 
$\Gamma(p\to \tau^+\to\bar{\nu}_{\tau}\pi^+)$ is straightforward using the phenomenological 
Lagrangian written in terms of baryon fields given in the
Appendix~\ref{ap:taudec}, plus the Standard Model electroweak charged-current interaction:
\begin{align}\label{eq:ptaunudecay}
\begin{aligned}
\Gamma(p\to \tau^+\to\bar{\nu}_{\tau}\pi^+) &= \frac{m_p\,G_F^2 f_\pi^2}{8\pi}\,
\frac{(m_p^2-m_\pi^2)^2}{(m_p^2-m_\tau^2)^2}\,
\big| \beta (C_{LL}^e)_{1113} \big|^2 \\[2mm]
&
\simeq (8.9\cdot 10^{-19}~\mbox{GeV}^5) \, \big|  (C_{LL}^e)_{1113} \big|^2
\,.
\end{aligned}
\end{align}
Given that the experimental bound on $p\to \bar{\nu}\pi^+$ is 
very strong, $\tau_{p\to\bar{\nu} \pi^+}> 25\times 10^{30}$ years~\cite{Beringer:1900zz}, we
can obtain  a  stringent limit for  $(C_{LL}^e)_{1113}$
from the virtual tau amplitude:
\begin{align}\label{eq:C1113bound}
\big|  (C_{LL}^e)_{1113} \big|< 3.1\cdot 10^{-17}~\mbox{TeV}^{-2}
\,.
\end{align}
According to this bound the possibility to observe any of the baryon number violating tau 
decays in Eq.~(\ref{eq:taudecaysnum}) seems to be far beyond the reach of future experiments,
an observation that already was noticed by Marciano~\cite{Marciano:1994bg} some time ago.
\par
Let us finally comment on the constraints on the coefficients $(C_{LL}^\nu)_{111i}$, for $i=1,2,3$
families, which are
also generated in our model. These can be obtained from the direct decay of the proton
into a pion and an anti-neutrino. The partial decay width formula for this process is similar to 
that in Eq.~(\ref{eq:pdecay}), and reads~\cite{Nath:2006ut}:
\begin{align}\label{eq:pnudecay}
\begin{aligned}
\Gamma(p\to \bar{\nu}_i\pi^+) &= \frac{(m_p^2-m_\pi^2)^2}{64\pi f_\pi^2\, m_p^3}\,
 \big| \beta (C_{LL}^\nu)_{111i} \big|^2 \, (1+D+F)^2 
\\[2mm]
&
\simeq (3.9\cdot 10^{-4}~\mbox{GeV}^5) \, \big|  (C_{LL}^\nu)_{111i} \big|^2
\,,
\end{aligned}
\end{align}
which implies 
\begin{align}
\big|  (C_{LL}^\nu)_{111i} \big| < 1.5\cdot 10^{-24}~\mbox{TeV}^{-2}
\,,
\label{eq:C111ibound}
\end{align}
using the experimental limit on $\tau_{p\to\bar{\nu} \pi^+}$ already mentioned.

\section{Conclusions}
\label{sec:conclusions}
Processes with $\Delta B = \Delta L \neq 0$ are allowed in the SM through quantum corrections
generated by instanton solutions of the Yang-Mills theory. However, when computed, these 
transitions turn out to be negligible. This is due to the smallness of the gauge coupling $g$.
In gauge extensions of the SM this suppression is still present but could be reduced for 
higher values of the coupling (and still small enough to allow for a perturbative treatment).
\par
We have presented a detailed analysis of the dynamics of $\Delta B = \Delta L = 1$ processes
generated by instantons corresponding to a gauge-extended model that breaks universality in 
the family couplings, the Non-Universal $G(221)$ model. We have determined the associated 
fermion zero modes, which are the main tool for the construction of the instanton-induced 
effective operator that generates those processes, and we have detailed the latter. Within 
a slight different framework this had already been studied in Ref.~\cite{Morrissey:2005uza} 
but there the inter-family mixing was not taken into account. 
\par
Once the effective action has been constructed we have proceeded to analyse proton decay in 
this framework, together with correlated tau decays into baryons (plus mesons or leptons). 
As expected the strong bound on the decay of the proton dominates clearly the information
on the couplings of the theory. Moreover it pushes any $\Delta B = \Delta L=1$ tau decay 
beyond the reach of any foreseen facility. However this should not discourage the experimental
hunt for those processes, as recently carried out by LHCb \cite{Aaij:2013fia}, because 
we still do not know which features nature prefers to extend the SM. In particular Belle II,
or super-B factories in general, could provide an appropriate setting to hunt for those 
decays of the tau lepton.
\par
Though we have carried out our study in a particular extension of the SM, the only relevant
feature is the fact that only one family couples to the Yang-Mills group whose instantons
are considered in the generation of the interaction. Our results can be extended straightforwardly
to any model with that property.

\section*{Acknowledgements}
This research has been supported in part by the Spanish Government, Generalitat Valenciana and
ERDF funds from the EU Commission [grants FPA2011-23778, PROMETEOII/2013/007, CSD2007-00042 
(Consolider Project CPAN)]. J. F. also acknowledges VLC-CAMPUS for an 
``Atracci\'{o} del Talent"  scholarship.

\appendix
\renewcommand{\theequation}{\Alph{section}.\arabic{equation}}
\renewcommand{\thetable}{\Alph{section}.\arabic{table}}

\section{One-instanton-generated fermion Green function}
\label{ap:GreenFunc}
\setcounter{equation}{0}
\setcounter{table}{0}

Let us consider a $SU(2)$ Yang-Mills theory with $A_{\mu}^a$ the gauge fields, a scalar sector provided by $\Phi_s$ Higgs 
representations with {\em isospin} $q_s$, $s=1,...N_S$, being $N_S$ the number of scalar multiplets and a matter content of
massless Weyl $\psi_i$ fermion doublets (both quarks and leptons), $i=1,..., N_f$. This system is described by the Euclidean effective action $S_{\mbox{\tiny E}}[\psi_i,A,\eta,\Phi_s]$ with $\eta$ the ghost fields.
We consider the vacuum to vacuum Green function that involves the fermion doublets with possible violation of flavour, lepton and baryon number and that is generated by the classical instanton solution that minimizes the Euclidean action $S_{\mbox{\tiny E}}$:
\begin{align}
\label{eq:GF}
G\left(x_1,\dots,x_{\mbox{\tiny $N_f$}}\right)=\langle\prod_{i=1}^{\mbox{\tiny $N_f$}}\psi_i\left(x_i\right)\rangle_{I} \, , 
\end{align}
where the sub index, $I$, stresses that the Green function is evaluated in an instanton background.
In the path integral formalism the evaluation of the Green
function requires an integration over the field configurations that are involved in our physical system. 
The method to perform the integration is based on a perturbative semi-classical expansion of the Euclidean action 
around the classical instanton configuration up to one-loop level using the  Background Field Method (BFM). In this expansion the fermion and ghost fields remain at the quantum level while both gauge fields and scalars are split into a classical instanton background field, labelled by $I$, and a quantum fluctuation:
\begin{align}\label{eq:bfmf}
\begin{aligned}
A_\mu^a &= A_\mu^{a,I}+A_\mu^{a,\mbox{\tiny q}}, \\
\Phi_s &= \Phi^I_s+\Phi_s^{\mbox{\tiny q}}.
\end{aligned}
\end{align}
\par
Hence within this setting the Green function takes the following form:
\begin{align}
\label{eq:GFpath}
G\left(x_1,\dots,x_{\mbox{\tiny $N_f$}}\right)=\frac{\int D\left[\psi_i,A,\eta,\Phi_s\right] \, e^{-S_{\mbox{\tiny E}}[\psi_i,A,\eta,\Phi_s]} \, \prod_{i=1}^{\mbox{\tiny $N_f$}}\psi_i\left(x_i\right)}{\int D\left[\psi_i,A,\eta,\Phi_s\right] \, e^{-S_{\mbox{\tiny E}}[\psi_i,A^{\mbox{\tiny q}},\eta,\Phi_s}]},
\end{align}
being $D\left[\psi_i,A,\eta,\Phi_s\right]$ the path integral measure. Notice that in the denominator the effective action does not depend on the instanton solutions as otherwise it would vanish because of the presence of zero modes (see below). 
Under the BFM expansion, the action now takes the form:
\begin{align}
\label{eq:effa}
\begin{aligned}
S_{\mbox{\tiny E}}\left[\psi_i,A,\eta,\Phi_s\right]=&\;S_{\mbox{\tiny E}}\left[A^I,\Phi^I_s\right]\, - \, \frac{1}{2}\, A^{\mbox{\tiny q}}\theta_A \, A^{\mbox{\tiny q}}\, + \, \overline{\psi_i}\, \theta_{\psi_i} \, \psi_i \, - \, \left(\Phi_s^{\mbox{\tiny q}}\right)^\dagger \, \theta_{\Phi_s} \, \Phi_s^{\mbox{\tiny q}} \, - \, \overline{\eta} \, \theta_\eta \, \eta\\
& + \mathcal{O}\left(A^{\mbox{\tiny q}},\psi_i,\Phi_s^{\mbox{\tiny q}},\eta\right)^3,
\end{aligned}
\end{align}
where we have abbreviated the interactions with the quantum fields by using $\theta_y$ for $y= A, \psi_i, \Phi_s, \eta$. In 
Eq.~(\ref{eq:effa}) $S_{\mbox{\tiny E}} \left[A^I, \Phi^I_s\right]$ is the action when only the background instanton fields are considered:
\begin{equation}
\label{eq:euclaction}
S_{\mbox{\tiny E}}\left[A,\Phi_s \right] \, = \, \int  d^4 x_{\mbox{\tiny E}} \, \left[ \frac{1}{2} \, \mbox{tr}\left[ F_{\mu \nu} F_{\mu \nu} \right] \, + \, 
\sum_s \mbox{Tr}\left[\left( D_{\mu} \Phi_s \right)^{\dagger} \left( D_{\mu} \Phi_s \right)\right]  \,  +  \,  V\left( \Phi_s \right) \, \right] ,
\end{equation}
where $F_{\mu \nu} = \partial_{\mu} A_{\nu} - \partial_{\nu} A_{\mu} - i g \left[ A_{\mu}, A_{\nu} \right]$ and $D_{\mu} = 
\partial_{\mu} - i g A_{\mu}$ with $A_{\mu} = A_{\mu}^a \tau^a/2$ and $\tau^a$ the Pauli matrices. 

One can perform Gaussian integration for the non-zero eigenvalues of the operators $\theta_y$ and compute the determinants coming from this integration by diagonalization.
However, one should take into account that the operators $\theta_y$ may have zero eigenvalues so one has to take care of the zero eigenfunction or zero modes before performing Gaussian integration. These zero modes can be taken into account by introducing the \textit{collective coordinates} formalism \cite{'tHooft:1976fv}. There are eight independent zero modes for the gauge boson related to the classical symmetries broken by the instanton solution: four translations, one dilatation and three global gauge transformations. In contrast, the operator $\theta_\eta$ contains no zero modes. Finally, the fermion operator $\theta_\psi$ also contains zero modes and they are treated in Section~\ref{sec:B+Llag}.
\par
Before proceeding let us comment on the fermion piece of the Green function in Eq.~(\ref{eq:GFpath}). This is given by the
generating functional:
\begin{equation} \label{eq:gffermion}
Z[\chi, \overline{\chi}] \, = \,  \int D[\psi] \, D[\overline{\psi}] \, \exp \left[ - \int d^4x \left( \, \overline{\psi} \,   \theta_\psi \, 
\psi \, - \,  \overline{\chi} \,  \psi \, - \,  \overline{\psi} \, \chi \,  \right) \right].
\end{equation}
$\chi(x)$ and $\overline{\chi}(x)$ are the external sources, anticommuting elements of an infinite-dimensional Grassmann algebra, and $\theta_\psi$ is, in general, a non-hermitian operator. Green functions are obtained by differentiating the generating functional with respect to $\chi$ and/or $\overline{\chi}$ with $\chi = \overline{\chi} = 0$. It can be shown (see for instance Ref.~\cite{Espinosa:1989qn}) that from the sector with no zero modes the only non-vanishing Green functions are those containing equal number of $\psi$ and $\overline{\psi}$ fields, subsequently conserving any fermion number. The fermion zero modes are the ones that generate a violation of fermion number as first noticed by 't Hooft \cite{'tHooft:1976fv,'tHooft:1976up}.
\par
After integrating over the field configurations, the Green function in terms of the \textit{collective coordinates} $y$, $\rho$ and $U$ takes the form:
\begin{align}
\label{eq:picc}
G\left(x_1,\dots,x_{\mbox{\tiny $N_f$}}\right)=\int d^4 y\int d\rho\int dU\,e^{-S_{\mbox{\tiny E}}\left[ A^I, \Phi^I_s \right]} \, F\left(\rho;\mu\right)\, \prod_{i=1}^{\mbox{\tiny $N_f$}}\psi_{0,i}\left(x_i-y\right),
\end{align}
where $\psi_{0,i}$ are the zero modes, located at $y$, associated to the fermion operator, $dU$ is the Haar measure of the $SU(2)$ instanton orientation and $\rho$ is the size of the instanton. In Eq.~(\ref{eq:picc}) $F\left(\rho;\mu\right)$ contains the contribution from the regularized product of non-zero eigenvalues ($\mu$ stands for the normalization point in the $\overline{\mbox{MS}}$ scheme) and other factors coming from the use of collective coordinates. This function was calculated in Ref.~\cite{'tHooft:1976fv}:
\begin{align}
\label{eq:frhomu}
F\left(\rho;\mu\right) \, = \, C \, g^{-8} \, \left(\rho\mu\right)^{\beta_1}\, \rho^{-5},
\end{align}
with $\beta_1$ being the $SU(2)$ one-loop beta function for our system:
\begin{align}
\label{eq:onelbeta}
\beta_1=\frac{22}{3}-\frac{1}{3}N_f-\frac{1}{6}N_S,
\end{align}
and $C$ is given by:
\begin{align}
\label{eq:cthooft}
C=2^{10}\pi^6e^{-\alpha\left(1\right)+\left(N_f-N_S\right)\alpha\left(\frac{1}{2}\right)+\frac{5}{36}\left(2-\frac{1}{2}N_f+\frac{1}{2}N_S\right)},
\end{align}
where $\alpha(1)\simeq0.443$ and $\alpha\left(\frac{1}{2}\right)\simeq0.146$.
\par
The momentum space Green function is finally:
\begin{align}\label{eq:GF2}
\widetilde{G}\left(p_1,\dots, p_{N_f}\right)= \left(2\pi\right)^4\delta^4\left(\sum_{i=1}^{N_f} p_i\right) \, \int dU\int d\rho\,e^{-S_{\mbox{\tiny E}}\left[ A^I, \Phi^I_s \right]}
F\left(\rho;\mu\right)\prod_{i=1}^{N_f} \widetilde{\psi}_{0,i}\left(p_i\right),
\end{align}
being $\widetilde{\psi}_{0,i}(p)$ the Fourier transform of the zero modes associated to the fermion operator. 
\par
Let us work now the remaining classical action of the background instanton fields $S_{\mbox{\tiny E}}\left[ A^I, \Phi^I_s \right]$. 
It is important to notice that  when $\langle\Phi_s^0\rangle\neq0$, the action has no non-trivial stationary points. 
An approximate instanton solution for small $\rho\langle\Phi_s^0\rangle$  that reduces to the classical solution for $\langle\Phi_s^0\rangle=0$, was anticipated by 't Hooft \cite{'tHooft:1976fv} and formally obtained by Affleck \cite{Affleck:1980mp} under the so-called \textit{constrained instanton formalism}.
As a result of this formalism, Affleck showed that while in the short-distance regime the instantonic solution behaves as in the case when $\langle\Phi_s^0\rangle=0$, in the long-distance regime it presents an exponential fall off. Using the \textit{singular gauge} for the gauge field, the constrained instanton solution reads:
\begin{align}
\label{eq:AI}
g \, A_\mu^{I \, a}\left(x\right)  \, \frac{\tau^a}{2} &=  \mathscr{A}^I(x)\, x_{\nu} \,  \overline{\eta}_{\mu\nu}^a \, U \, \tau^a \, U^\dagger \,,
\end{align}
with
\begin{align}\label{eq:Arara}
\mathscr{A}^I(x)= \begin{cases} 
                             \; \;   \rho^2 \,  \frac{1}{x^2 \, \left(x^2+\rho^2\right)}  \,, & x\ll\rho , \\[0.6cm]
                             \; \;    \rho^2 \, M_W^2 \, \frac{K_2\left(M_Wx\right)}{2 \, x^2}  \,, & x\gg\rho , \\
                                \end{cases}
\end{align}
where $M_W$ is the mass of the gauge boson generated by the spontaneous symmetry breaking of the Higgs sector.
The corresponding anti-instanton solution is given by $\bar{A}_{\mu}^{I} = A_{\mu}^{I} \left( \overline{\eta}_{\mu \nu}^a \longrightarrow \eta_{\mu \nu}^a \right)$, where $\eta_{\mu\nu}^a$ and $\overline{\eta}_{\mu\nu}^a$ are the t' Hooft symbol and its self-dual which relate the $SO(4)$ generators to the $SU(2)$ generators (see Ref.~\cite{'tHooft:1976fv}). 
\par
The instanton solution for the Higgs field is:
\begin{align}
\label{eq:PhiI}
\Phi_s^I\left(x\right)&=   \begin{cases}
                              \; \;   \left(\frac{x^2}{x^2+\rho^2}\right)^{q_s} \, \langle\Phi_s^0\rangle \, \overline{h}, & x\ll\rho , \\[0.6cm]
                              \; \;   \left( 1 \,  - \, \rho^2 \, M_H \, \frac{K_1\left(M_Hx\right)}{2 \, x} \right) \, \langle \Phi_s^0 \rangle \,  \overline{h}, & x\gg\rho , \\
                                \end{cases}
\end{align}
where $M_H$ is the Higgs boson mass after spontaneous symmetry breaking, $\overline{h}=\left(0,1\right)^T$ is a constant isospinor, $q_s$ is the \textit{isospin} of the scalar under $SU(2)_h$ and  the $K_{\nu}(x)$ are Modified Bessel functions of the second kind.
\par
The action for the constrained instanton can be calculated, perturbatively in $X_s \equiv \rho \,  \langle \Phi_s^0 \rangle$, 
using the solutions above for $A_{\mu}^{I}$ and $\Phi_s^I$, giving:
\begin{align}
\label{eq:eaction}
\begin{aligned}
\frac{1}{2}\int d^4 x_{\mbox{\tiny E}} \, \mbox{Tr} \left[F_{\mu \nu} F_{\mu \nu} \right]  & =  \frac{8\pi^2}{g^2} \, + \, \mathcal{O}\left[X_s^4\right], \\
\int d^4 x_{\mbox{\tiny E}} \; \mbox{Tr}\left[\left|D_\mu\Phi_s\right|^2\right] & =  4  \, \pi^2 \, q_s \, X_s^2 \, + \, \mathcal{O}\left[X_s^4\ln X_s\right], \\
\int d^4 x_{\mbox{\tiny E}}\, V\left(\Phi_s\right)& =  \mathcal{O}\left[ X_s^4\ln X_s \right],
\end{aligned}
\end{align}
and therefore the leading contribution is given by:
\begin{equation}
\label{eq:SCL}
S_{\mbox{\tiny E}}\left[A^I,\Phi^I_s\right] \, \simeq \, \frac{8\pi^2}{g^2} \, + \, 4 \, \pi^2 \rho^2 \, \, \mathcal{V}^2,
\end{equation}
with $\mathcal{V}^2=\sum_s \, q_s \, \langle \Phi_s^0 \rangle^2$. The action provided by the instanton field and its corresponding anti-instanton is the same. 
As shown in Ref.~\cite{'tHooft:1976fv}, the factor coming from the constrained scalar fields ensures the convergence of the integral over instanton size in the infra-red regime, that is for $\rho\to\infty$.

\section{Instanton solutions in the \texorpdfstring{$\boldsymbol{G(221)}$}{G(221)} model}
\label{ap:mixing}
\setcounter{equation}{0}
\setcounter{table}{0}

In this section we intend to apply the derivation done in Appendix~\ref{ap:GreenFunc} to the Non-Universal $G(221)$ model case. One could think that the computation of the instanton field configurations in frameworks with more complicated gauge structures could give rise to troublesome mixing effects among the instanton solutions for the different gauge fields. We will see that, at first order, this is not the case and a separate treatment of the instanton configurations is possible. 

The equations of motion for the background fields (see Eqs.~\eqref{eq:bfmf}) in the Non-Universal $G(221)$ model are given by:
\begin{align}\label{eq:G221_EOM}
\begin{aligned}
D_\mu W_{h\,\mu\nu}^a&=-\frac{1}{2}ig_h\left[\Phi_h^\dagger\tau^a D_\mu \Phi_h-\left(D_\mu\Phi_h\right)^\dagger\tau^a\Phi_h\right]+\frac{1}{2}ig_h\left[\Phi_b^\dagger\tau^a D_\mu \Phi_b-\left(D_\mu\Phi_b\right)^\dagger\tau^a\Phi_b\right],\\
D_\mu W_{l\,\mu\nu}^a&=-\frac{1}{2}ig_l\left[\Phi_l^\dagger\tau^a D_\mu \Phi_l-\left(D_\mu\Phi_l\right)^\dagger\tau^a\Phi_l\right]-\frac{1}{2}ig_l\left[\Phi_b^\dagger\tau^a D_\mu \Phi_b-\left(D_\mu\Phi_b\right)^\dagger\tau^a\Phi_b\right],\\
\partial_\mu B_{\mu\nu}&=-ig^\prime\left[\Phi_h^\dagger D_\mu \Phi_h-\left(D_\mu\Phi_h\right)^\dagger\Phi_h\right]-ig^\prime\left[\Phi_l^\dagger D_\mu \Phi_l-\left(D_\mu\Phi_l\right)^\dagger\Phi_l\right],\\
D^2\Phi_i&=\frac{\delta V\left(\Phi_s\right)}{\delta \Phi_i}.
\end{aligned}
\end{align}
These equations suggest that the classical solutions of $W_l$ and $W_h$ are mixed and therefore, given a non-zero classical field configuration for the $SU(2)_h$ gauge field, $W^{cl}_h$, a mixing will be induced through the bi-doublet giving rise to a non-zero $W^{cl}_l$. However it is well known that, in the presence of scalars that acquire a vacuum expectation value, there is no classical solution of the action. One can obtain an approximate solution that resembles the instantonic one in the region where $\rho\langle\Phi_s\rangle\ll1$ with an exponential fall off outside this region. This is achieved with the use of the {\em constrained instanton formalism}~\cite{Affleck:1980mp}, which relies in a formal expansion in $\rho\langle\Phi_s\rangle$ to obtain this solution. At first order in this expansion the rhs terms in Eqs.~\eqref{eq:G221_EOM} can be neglected and the equations of motion now read, up to $\mathcal{O}\left(\rho^2\langle\Phi_s\rangle^2\right)$:
\begin{align}\label{eq:G221_EOM_constrained}
\begin{aligned}
D_\mu W_{h\,\mu\nu}^a&=0,\\
D_\mu W_{l\,\mu\nu}^a&=0,\\
\partial_\mu B_{\mu\nu}&=0,\\
D^2\Phi_i&=0,
\end{aligned}
\end{align}
such that they reduce to decoupled instanton equations for each gauge group. Several instanton solutions to Eqs.~\eqref{eq:G221_EOM_constrained} are possible and they will induce baryon and lepton number violating Green functions of different nature:
\begin{itemize}
 \item[i)] $W_l^{cl}=W_l^I$, $W_h^{cl}=W_h^I$ mediating processes with $\Delta B=\Delta L=3$,
 \item[ii)] $W_l^{cl}=W_l^I$, $W_h^{cl}=0$ mediating processes with $\Delta B=\Delta L=2$,
 \item[iii)] $W_l^{cl}=0$, $W_h^{cl}=W_h^I$ mediating processes with $\Delta B=\Delta L=1$.
\end{itemize}

A difficulty appears if we want to consider physical gauge fields as asymptotic states. In this case we should include the mixing introduced by the scalar fields, which forces us to consider the next order in the constrained instanton expansion. In this way one would generate classical field configurations that are no longer $W_l^I$ and $W_h^I$ but $W_I$ and $W^\prime_I$ and the instanton solutions are mixed. This approach was followed by Gibbs~\cite{Gibbs:1995xt} in the SM case. As we are dealing with instanton Green functions that do not contain any external gauge boson, this is irrelevant to our computation and we can safely work in the first order approximation. 

Finally, it is interesting to remark that although the derivations done in this appendix have focused in the $G(221)$ model case they can be easily extrapolated to other frameworks. In particular, one straightforward result of this discussion is that for the SM-instanton Green functions with no gauge fields as asymptotic states the effects of a non-zero $\theta_W$ are not relevant at first order in $\rho\langle\Phi\rangle$, just like in the case considered here.

\section{The projector operators}
\label{ap:projectors}
\setcounter{equation}{0}
\setcounter{table}{0}

In this section we define the projector operators in terms of the unitary matrices which relate the gauge-diagonal basis and the mass-diagonal basis and explicitly show that these operators are projectors. The states from the different basis are related in the following way:
\begin{align}\label{eq:mass-basis}
\begin{aligned}
u^{\left(ph\right)}_i=\left(V_u\right)_{ij}u_j^{\left(g\right)},  \\
d^{\left(ph\right)}_i=\left(V_d\right)_{ij}d_j^{\left(g\right)},  \\
\nu^{\left(ph\right)}_i=\left(V_\ell\right)_{ij}\nu_j^{\left(g\right)}, \\
e^{\left(ph\right)}_i=\left(V_\ell\right)_{ij}e_j^{\left(g\right)},
\end{aligned}
\end{align}
where we have used the $\left(ph\right)$ and $\left(g\right)$ superscripts to denote states in the mass-diagonal and gauge-diagonal basis, respectively.
\par
Using the family rotations in Eq.~(\ref{eq:mass-basis}) we define the projection over the third family as:

\begin{align}\label{eq:pdss}
\begin{aligned}
\mathcal{P}_u\,u^{\left(ph\right)}_i&=\left(V_u\right)_{i3}u_3^{\left(g\right)}  =\left(V_u\right)_{i3}\left(V_u^\dagger\right)_{3j}u^{\left(ph\right)}_j,  \\
\mathcal{P}_d\,d^{\left(ph\right)}_i&=\left(V_d\right)_{i3}d_3^{\left(g\right)}  =\left(V_d\right)_{i3}\left(V_d^\dagger\right)_{3j}d^{\left(ph\right)}_j,  \\
\mathcal{P}_\ell\,\nu^{\left(ph\right)}_i&=\left(V_\ell\right)_{i3}\nu_3^{\left(g\right)}  =\left(V_\ell\right)_{i3}\left(V_\ell^\dagger\right)_{3j}\nu^{\left(ph\right)}_j,  \\
\mathcal{P}_\ell\,e^{\left(ph\right)}_i&=\left(V_\ell\right)_{i3}e_3^{\left(g\right)}  =\left(V_\ell\right)_{i3}\left(V_\ell^\dagger\right)_{3j}e^{\left(ph\right)}_j.
\end{aligned}
\end{align}

This allow us to define the projector operators $P_f$ with $f=u,\,d,\,l$ as:
\begin{align}
\begin{aligned}
\left(P_f\right)_{ij}&=\left(V_f\right)_{i3}\left(V_f^\dagger\right)_{3j}.
\end{aligned}
\end{align}
The operators defined this way present several interesting properties:
\begin{enumerate}[(i)]
 \item They are self-adjoint, as can be trivially seen from their definition.
 \item\label{enum:idempotence} They are projector operators. This can be proved by showing their idempotence:
 \begin{align} \label{eq:puupex}
 \begin{aligned}
  \left(P_f^2\right)_{ij}&=\left(V_f\right)_{i3}\left(V_f^\dagger\right)_{3k}\left(V_f\right)_{k3}\left(V_f^\dagger\right)_{3j}\\
  &=\left(V_f\right)_{i3}\left(V_f^\dagger\right)_{3j}\\
  &=\left(P_f\right)_{ij},
 \end{aligned}
 \end{align}
 where we have used the unitarity of the $V_f$ matrix.
 \item\label{enum:rank} The projectors have rank one. This becomes obvious after we write the operators in matrix form,
 \begin{align}
 P_f=
 \begin{pmatrix}
 \left|V_f\right|^2_{13} & \left(V_f\right)_{13}\left(V_f\right)_{23}^* & \left(V_f\right)_{13}\left(V_f\right)_{33}^*\\
 \left(V_f\right)_{23}\left(V_f\right)_{13}^* & \left|V_f\right|^2_{23} & \left(V_f\right)_{23}\left(V_f\right)_{33}^*\\
 \left(V_f\right)_{33}\left(V_f\right)_{13}^* & \left(V_f\right)_{33}\left(V_f\right)_{23}^* & \left|V_f\right|^2_{33}
 \end{pmatrix},
 \end{align}
and notice that all columns are proportional to the vector
$((V_f)_{13},(V_f)_{23}, (V_f)_{33})$.
 
 \item\label{enum:projection} As consequence of properties (\ref{enum:idempotence}) and (\ref{enum:rank}), these operators will project any vector in flavour space into a one-dimensional subspace. In this case the subspace is given by:
 \begin{align}
 \left\{\alpha\left[\left(V_f\right)_{13}u_{1}^{(ph)}+\left(V_f\right)_{23}u_{2}^{(ph)}+\left(V_f\right)_{33}u_{3}^{(ph)}\right]\;:\;\alpha\in\mathds{C}\right\}. 
 \end{align}
 
 \item\label{enum:relation} The up- and down-type projectors in the quark sector are related through the CKM matrix in the following way:
 \begin{align}\label{eq:relation}
 \begin{aligned}
 P_u&=V_{{\mbox{\tiny CKM}}}\,P_d\,V_{{\mbox{\tiny CKM}}}^\dagger,\\
 P_u&=V_{{\mbox{\tiny CKM}}}^\dagger\,P_d\,V_{{\mbox{\tiny CKM}}}.
 \end{aligned}
 \end{align}
 This is readily proven:
 \begin{align}
 \begin{aligned}
 \left(V_{{\mbox{\tiny CKM}}}\,P_d\,V_{{\mbox{\tiny CKM}}}^\dagger\right)_{ij}&=\left(V_u\right)_{im}\left(V_d^\dagger\right)_{mn}\left(V_d\right)_{n3}\left(V_d^\dagger\right)_{3k}\left(V_d\right)_{kl}\left(V_u^\dagger\right)_{lj}\\
 &=\left(V_u\right)_{i3}\left(V_u^\dagger\right)_{3j}\\
 &=\left(P_u\right)_{ij},
 \end{aligned}
 \end{align}
 with the CKM matrix defined as $V_{{\mbox{\tiny CKM}}}=V_u\,V_d^\dagger$ and where we have used that the matrices $V_f$ are unitary. The second identity can be obtained from the first one by hermitic conjugation and using that the projectors are self-hermitian.
\end{enumerate}

\section{The integration over instanton gauge orientation}
\label{ap:gaugeInt}
\setcounter{equation}{0}
\setcounter{table}{0}

In this Appendix we will perform the integration over the instanton gauge orientation ($U$) in the amplitude in Eq.~(\ref{eq:amplitude1}).  In order to simplify the discussion we will omit the flavour indices in this Appendix.
We follow closely Ref.~\cite{Morrissey:2005uza}. To proceed we parametrize the $SU(2)$ group elements in terms of the Pauli matrices such that
$U=e^{i\alpha\hat{n}\cdot\vec{\sigma}}$ with $\hat{n}=\left(\sin\theta\cos\phi,\sin\theta\sin\phi,\cos\theta\right)$ is a unit vector.
The coordinate ranges are:
$\alpha\in\left[0,2\pi\right]$, $\theta\in\left[0,\pi\right]$ and $\phi\in\left[0,2\pi\right]$.
For this parametrization the Haar measure takes the form:
\begin{align}
\int dU=\frac{1}{4\pi^2}\int_0^{2\pi} d\alpha\sin^2\frac{\alpha}{2}\int_{-1}^1d\cos\theta\int_0^{2\pi}d\phi.
\end{align}
\par
In our particular case, $N_f=4$, we have a product of four U matrices: each up-type fermion introduces a factor 
$\overline{\omega}_{u}  \, U \, \chi_u \, = \, \overline{\omega}_{u} \, \left(U_{12},U_{22}\right)$ while each down-type fermion gives a factor $\overline{\omega}_{d}   \, U \, \chi_d \, = \, \overline{\omega}_{d} \, \left(-U_{11},-U_{21}\right)$ and the same for the leptonic sector. From the group integration of the product of the four matrices, only three combinations survive:
\begin{align} \label{eq:duu}
\begin{aligned}
&\int dU\, U_{11}^2U_{22}^2=\frac{1}{3}, \\
&\int dU\, U_{12}^2U_{21}^2=\frac{1}{3}, \\
&\int dU\, U_{11}U_{22}U_{12}U_{21}=-\frac{1}{6}.
\end{aligned}
\end{align}
Now, introducing $SU(3)_c$ indices we find the structures:
\begin{align}\label{eq:Uterms}
\begin{aligned}
&u^1 \, u^2 \, d^3 \, e, \qquad u^1 \, d^2 \, u^3 \, e, \qquad d^1 \, u^2 \, u^3 \, e,\\
&d^1 \, d^2 \, u^3 \, \nu, \qquad d^1 \, u^2 \, d^3 \, \nu, \qquad u^1 \, d^2 \, d^3 \, \nu.
\end{aligned}
\end{align}
All these terms have the same sign and the same prefactor.
\par
Let us consider, for instance, the first structure in Eq.~(\ref{eq:Uterms}). After performing the $SU(2)$ integration, the terms contributing to this structure are:
\begin{align}\label{eq:spinorus}
\begin{aligned}
\int dU & \, \overline{\omega}_{u^1} \,  \left(U_{12},U_{22}\right) \, \overline{\omega}_{u^2} \,  
\left(U_{12},U_{22}\right) \, \overline{\omega}_{d^3} \, \left(-U_{11},-U_{21}\right) \, \overline{\omega}_e 
 \left(-U_{11},-U_{21}\right)\\
=& \, \frac{1}{3} \, \left[\overline{\omega}_{u^1,1}  \,  \overline{\omega}_{u^2,1} \,  
\overline{\omega}_{d^3,2} \,  \overline{\omega}_{e,2} \,  
+ \,  \overline{\omega}_{u^1,2}  \,  \overline{\omega}_{u^2,2} \,  
\overline{\omega}_{d^3,1} \, \overline{\omega}_{e,1} \, \right. \\
& \left. \; \; \; \; \;  - \, \frac{1}{2}\left(\overline{\omega}_{u^1,1} \, \overline{\omega}_{u^2,2} \, + \, \overline{\omega}_{u^1,2} \,  \overline{\omega}_{u^2,1} \right)
\left(\overline{\omega}_{d^3,2} \, \overline{\omega}_{e,2} \, + \, \overline{\omega}_{d^3,2} \, \overline{\omega}_{e,2} \right)\right],
\end{aligned}
\end{align}
where the second index in $\omega_{a,b}$ is spinorial. This spinor structure can be reproduced by the following effective operator:
\begin{align} \label{eq:efecoper}
\frac{1}{6}\overline{u^2_L}\left(d^3_L\right)^C\overline{u^1_L}\left(e_L\right)^C-\frac{1}{6}\overline{u^1_L}\left(d^3_L\right)^C\overline{u^2_L}\left(e_L\right)^C.
\end{align}
By writing explicitly the sum over colour indices,
\begin{align} \label{eq:scolor}
\epsilon_{\alpha\beta\gamma}u^\alpha u^\beta d^\gamma=2\left(u^1 \, u^2 \, d^3 \, + \, u^1 \, d^2 \, u^3 \, + \, d^1 \, u^2 \, u^3\right),
\end{align}
including all the terms coming from the integration over instanton gauge orientation and considering the flavour structure we get the result in Eq.~(\ref{eq:operatore}).

\section{\texorpdfstring{$\boldsymbol{\Delta B = \Delta L = 1}$}{BNV} tau decay rates}
\label{ap:taudec}
\setcounter{equation}{0}
\setcounter{table}{0}

We collect in this Appendix the results for several lepton and baryon number
violating tau decay rates with $\Delta B = \Delta L = 1 $
for which experimental bounds exist. The latter
include  hadronic decays with $\Delta S=0$ 
($\tau^+\to p\,\pi^0\,,p\,\eta\,,p\,\pi^0\pi^0,p\,\pi^0\,\eta$) and $\Delta S=1$ 
($\tau^+\to \Lambda\,\pi^+$), as well as electromagnetic decays
($\tau^+ \to p\,\gamma,\, p\,\mu^+\,\mu^-$).
\par
The rates provided are parametrized in terms of the  Wilson coefficients of all the baryon and
lepton number violating operators of dimension six involving the $\tau$ lepton~\cite{Weinberg:1979sa,Wilczek:1979hc}:
\begin{align}\label{eq:L_BL}
\begin{aligned}
\mathscr{L}_{B+L} \ = \  \frac{1}{\Lambda^2} \, \bigg[ & C_{RL}\,O_{RL}+ C_{LR}\,O_{LR}
+  C_{LL}\,O_{LL}+ C_{RR}\,O_{RR}\\[0mm]
& + \tilde{C}_{RL}\,\tilde{O}_{RL}+ \tilde{C}_{LR}\,\tilde{O}_{LR}
+  \tilde{C}_{LL}\,\tilde{O}_{LL}+ \tilde{C}_{RR}\,\tilde{O}_{RR}
\,  \bigg] \, + {\rm h.c.} \,,
\end{aligned}
\end{align}
where the operators without tilde generate interactions with $\Delta S=0$,
\begin{align}\label{eq:op}
\begin{aligned}
\mathcal{O}_{RL}&=\epsilon_{\alpha\beta\gamma}\overline{\left(u_R^{\alpha}\right)^C}d_R^{\beta}\overline{\left(u_L^{\gamma}\right)^C}\tau_{L} \,,\\
\mathcal{O}_{LR}&=\epsilon_{\alpha\beta\gamma}\overline{\left(u_L^{\alpha}\right)^C}d_L^{\beta}\overline{\left(u_R^{\gamma}\right)^C}\tau_{R} \,,\\
\mathcal{O}_{LL}&=\epsilon_{\alpha\beta\gamma}\overline{\left(u_L^{\alpha}\right)^C}d_L^{\beta}\overline{\left(u_L^{\gamma}\right)^C}\tau_{L} \,,\\
\mathcal{O}_{RR}&=\epsilon_{\alpha\beta\gamma}\overline{\left(u_R^{\alpha}\right)^C}d_R^{\beta}\overline{\left(u_R^{\gamma}\right)^C}\tau_{R} \,,
\end{aligned}
\end{align}
while operators with tilde contain the strange quark and produce interactions with $|\Delta S|=1$,
\begin{align}\label{eq:optilde}
\begin{aligned}
\tilde{\mathcal{O}}_{RL}&=\epsilon_{\alpha\beta\gamma}\overline{\left(u_R^{\alpha}\right)^C}s_R^{\beta}\overline{\left(u_L^{\gamma}\right)^C}\tau_{L} \,,\\
\tilde{\mathcal{O}}_{LR}&=\epsilon_{\alpha\beta\gamma}\overline{\left(u_L^{\alpha}\right)^C}s_L^{\beta}\overline{\left(u_R^{\gamma}\right)^C}\tau_{R} \,,\\
\tilde{\mathcal{O}}_{LL}&=\epsilon_{\alpha\beta\gamma}\overline{\left(u_L^{\alpha}\right)^C}s_L^{\beta}\overline{\left(u_L^{\gamma}\right)^C}\tau_{L} \,,\\
\tilde{\mathcal{O}}_{RR}&=\epsilon_{\alpha\beta\gamma}\overline{\left(u_R^{\alpha}\right)^C}s_R^{\beta}\overline{\left(u_R^{\gamma}\right)^C}\tau_{R} \,.
\end{aligned}
\end{align}
Here $\alpha,\,\beta$ and $\gamma$ are colour indices. 
Charge conjugation of the spinor fields is defined as usual, $\psi^C\equiv C \overline{\psi}^T$, which implies $\overline{\psi^C}= \psi^T C$.
Note that we have factored out the dependence on the scale of new physics from the Wilson coefficients in Eq.~(\ref{eq:L_BL}),
unlike in Eq.~(\ref{eq:lblcano}); the operator $\mathcal{O}_{LL}$ above is equal to $(\mathcal{O}_{LL}^e)_{1113}$ of Eq.~(\ref{eq:operatorae}), and
the corresponding Wilson coefficients are related by $C_{LL} / \Lambda^2=(C^{e}_{LL})_{1113}$.
Readers not interested in the details of the calculation
can skip to Eq.~(\ref{eq:rates}) and Tables~\ref{tab:rates} and~\ref{tab:rateseta}, which provide the result for the rates in terms
of the Wilson coefficients that parametrize the baryon and lepton number violating 
interactions.
\par
The transformation properties of the operators in Eqs.~(\ref{eq:op}) and (\ref{eq:optilde})  under  $G=SU(3)_L\otimes SU(3)_R$
and parity can be used to write the interactions in terms of meson and baryon fields~\cite{Claudson:1981gh,Nath:2006ut}.
To lowest order in derivatives the hadronic operators read:
\begin{align}\label{eq:oph}
\begin{aligned}
\mathcal{O}^{had}_{RL}&=\alpha\overline{\left(\tau_{L}\right)^C}\langle Pu^\dagger B_Lu^\dagger\rangle,\\
\mathcal{O}^{had}_{LR}&=- \, \alpha\overline{\left(\tau_{R}\right)^C}\langle PuB_Ru\rangle,\\
\mathcal{O}^{had}_{LL}&=\beta\overline{\left(\tau_{L}\right)^C}\langle Pu^\dagger B_Lu\rangle,\\
\mathcal{O}^{had}_{RR}&=- \, \beta\overline{\left(\tau_{R}\right)^C}\langle PuB_Ru^\dagger\rangle,\\
\tilde{\mathcal{O}}^{had}_{RL}&=\gamma\overline{\left(\tau_{L}\right)^C}\langle \tilde{P}u^\dagger B_Lu^\dagger\rangle,\\
\tilde{\mathcal{O}}^{had}_{LR}&=- \, \gamma\overline{\left(\tau_{R}\right)^C}\langle \tilde{P}uB_Ru\rangle,\\
\tilde{\mathcal{O}}^{had}_{LL}&=\delta\overline{\left(\tau_{L}\right)^C}\langle \tilde{P}u^\dagger B_Lu\rangle,\\
\tilde{\mathcal{O}}^{had}_{RR}&=- \, \delta\overline{\left(\tau_{R}\right)^C}\langle \tilde{P}uB_Ru^\dagger\rangle\;,
\end{aligned}
\end{align}
where: 
\begin{align}
P=\left(\begin{matrix}
0 & 0 & 0\\
0 & 0 & 0\\
1 & 0 & 0\\
\end{matrix}\right)\,, \quad\quad
\tilde{P}=-\left(\begin{matrix}
0 & 0 & 0\\
1 & 0 & 0\\
0 & 0 & 0\\
\end{matrix}\right) 
\,,
\end{align}
project out the $Q=+1$, $S=0$ and $Q=+1$, $S=+1$ hadronic components, respectively.
The unitary matrix $u(\phi)$ collects the Goldstone fields:
\begin{align}\label{eq:u}
u\left(\phi\right)=e^{-\frac{i}{\sqrt{2}f_0}\phi}\,, \quad  \quad 
\phi=\frac{1}{\sqrt{2}}\sum_{i=1}^8\lambda_i\phi_i=
\left(\begin{matrix}
\frac{\pi^0}{\sqrt{2}}+\frac{\eta_8}{\sqrt{6}} & \pi^+ & K^+\\
\pi^- & -\frac{\pi^0}{\sqrt{2}}+\frac{\eta_8}{\sqrt{6}} & K^0\\
 K^- & K^0 & -\frac{2\eta_8}{\sqrt{6}}\\
\end{matrix}\right)\,,
\end{align}
with $f_0$ being related to the pion decay constant, $f_0\simeq f_\pi\simeq 92.4$ MeV,
whereas baryons are introduced through the
$SU(3)$ matrix:
\begin{align}\label{eq:Bmultiplet}
B = 
\left( \begin{array}{ccc}
{\frac{1}{\sqrt{2}}}\Sigma^0 + {\frac{1}{\sqrt{6}}}\Lambda  & \Sigma^+  & p \\ 
\Sigma^- & - {\frac{1}{\sqrt{2}}}\Sigma^0 + 
{\frac{1}{\sqrt{6}}}\Lambda_{8} & n \\ 
\Xi^- & \Xi^{0} 
& -{\frac{2}{\sqrt{6}}}\Lambda 
\end{array}\right)
\,.
\end{align}
Using the transformation properties of the matrices $u(\phi)$ and $B$ for a given  $g=\left(g_R,g_L\right)\in G$,
\begin{align}
u\left(\phi\right)\stackrel{G}{\to}g_Ru\left(\phi\right)h\left(g,\phi\right)^{-1}=h\left(g,\phi\right)u\left(\phi\right)g_L^{-1}\,,
\quad  \qquad 
B\stackrel{G}{\to}h\left(g,\phi\right)B h\left(g,\phi\right)^{-1}
\,,
\label{eq:Gtrans}
\end{align}
where $h(g,\phi)$ is a compensating $SU(3)_V$ matrix, 
it is straightforward to check that the hadronic operators $O_{X}^{had}$ transform similarly to the partonic ones.
The strong coefficients $\alpha$, $\beta$, $\gamma$ and $\delta$ can be related to the matrix elements
of the three-quark field operators in Eq.~(\ref{eq:oph})  between a nucleon and the vacuum state:
\begin{align}\label{eq:strongpar}
\begin{aligned}
\langle0\mid \epsilon_{\alpha\beta\gamma}\overline{\left(u_R^{\alpha}\right)^C}d_R^{\beta}\overline{\left(u_L^{\gamma}\right)^C} 
\mid p(\mathbf{k})\rangle &=\alpha\, P_L\, u_p(\mathbf{k}) \;, \\
\langle0\mid\epsilon_{\alpha\beta\gamma}\overline{\left(u_L^{\alpha}\right)^C}d_L^{\beta}\overline{\left(u_L^{\gamma}\right)^C}
\mid p(\mathbf{k})\rangle &=\beta\, P_L\, u_p(\mathbf{k}) \;,\\
\langle0\mid\epsilon_{\alpha\beta\gamma}\overline{\left(u_R^{\alpha}\right)^C}s_R^{\beta}\overline{\left(u_L^{\gamma}\right)^C}
\mid \Sigma^+(\mathbf{k})\rangle &=\gamma\, P_L\, u_{\Sigma^+}(\mathbf{k}) \;,\\
\langle0\mid\epsilon_{\alpha\beta\gamma}\overline{\left(u_L^{\alpha}\right)^C}s_L^{\beta}\overline{\left(u_L^{\gamma}\right)^C}
\mid \Sigma^+(\mathbf{k})\rangle &=\delta\, P_L\, u_{\Sigma^+}(\mathbf{k}) \,,
\end{aligned}
\end{align}
where $u(\mathbf{k})$ is the spinor wave-function associated with the corresponding baryon of momentum $\mathbf{k}$.
Parity relates the matrix elements shown above
with those arising from the quark structure of operators $\mathcal{O}_{LR},\,\mathcal{O}_{RR}$  
and $\tilde{\mathcal{O}}_{LR},\,\tilde{\mathcal{O}}_{RR}$; for instance $\langle0\mid \mathcal{O}_{LR}\mid p(\mathbf{k})\rangle=-\alpha\, P_R\, u_p(\mathbf{k})$. Moreover, $SU(3)_V$ symmetry 
establishes that $\gamma=\alpha$ and $\delta=\beta$. 
This is explicitly tested in Ref.~\cite{Donoghue:1982jm} where the parameters where calculated under some simplifications. 
Parameters $\alpha$ and $\beta$ are known to satisfy the constraint $|\alpha|\simeq|\beta|$~\cite{Brodsky:1983st}. 
A lattice computation of parameters $\alpha$ and $\beta$ at the scale $Q=2\,\mbox{GeV}$
by the RBC-UKQCD collaboration
gives~\cite{Aoki:2008ku}: 
\begin{align}\label{eq:strongparnum}
\begin{aligned}
\alpha&=-0.0112(25)\mbox{ GeV}^3,\\
\beta&=0.0120(26)\mbox{ GeV}^3
\,,
\end{aligned}
\end{align}
where the phase convention has been chosen in such a way that the parameters $\alpha$ and $\beta$ are real.
\par
To compute the tau decay rates at tree-level, we need the lepton and baryon
number violating interaction vertices  $\tau \to p +n\phi$ with $n=0,1,2$ meson fields,
which are obtained by series expansion of the hadronic operators $O_{X}^{had}$ in $\phi$. 
In addition, the interactions that conserve baryon number enter the amplitudes with virtual hadrons. 
The latter are contained in the $SU(3)_L \otimes SU(3)_R$ invariant Lagrangian: 
\begin{equation}                                                                           
\mathscr{L}_{B}=   
\langle \,\overline{B} ( i\slashed{\nabla} -M_B) B \,\rangle
-{\frac{D}{2}}\,\langle \,\overline{B} \gamma^{\mu}\gamma_5 \{ u_{\mu},B \} \, \rangle  
-{\frac{F}{2}}\,\langle \,\overline{B} \gamma^{\mu}\gamma_5 [ u_{\mu},B ] \, \rangle + \dots \,,
\label{eq:L_B}                                                                                               
\end{equation}
where the dots stand for terms with more derivatives.
The covariant derivative 
\begin{equation}   
\nabla_{\mu}B=\partial_{\mu}B+[\Gamma_{\mu},V]\,,\ \ \ \  \ \ \ \ 
\Gamma_{\mu}=\frac{1}{2}\left[ u^{\dagger}(\partial_{\mu} - i r_{\mu})u+
u(\partial_{\mu} - i \ell_{\mu})u^{\dagger}\, \right] , 
\label{eq:covder}                                                                                               
\end{equation}
is defined in such a way that $\nabla_{\mu}B$ transforms in the same way as the
baryon matrix $B$, Eq.~(\ref{eq:Gtrans}). 
The left and right source fields, $l_\mu$, $r_\mu$, reproduce the couplings of the 
baryons to  external vector and axial-vector currents, and 
$u_\mu$ is the chiral tensor familiar
from $\chi$PT:
\begin{align}
u_\mu=i\left[u^\dagger\left(\partial_\mu-ir_\mu\right)u-u\left(\partial_\mu-il_\mu\right)u^\dagger\right]
\,.
\end{align}
For the constants $F$ and $D$ in Eq.~(\ref{eq:L_B}),
we shall use the values: 
\begin{equation} 
F +D = 1.2670 \pm 0.0030, \quad\quad  F-D = -0.341 \pm 0.016
\,,
\label{eq:FDnum}
\end{equation}
obtained from an analysis of hyperon decays in Ref.~\cite{Cabibbo:2003cu}.   
\begin{figure}[t!]
\begin{centering}
\includegraphics[width=1\textwidth]{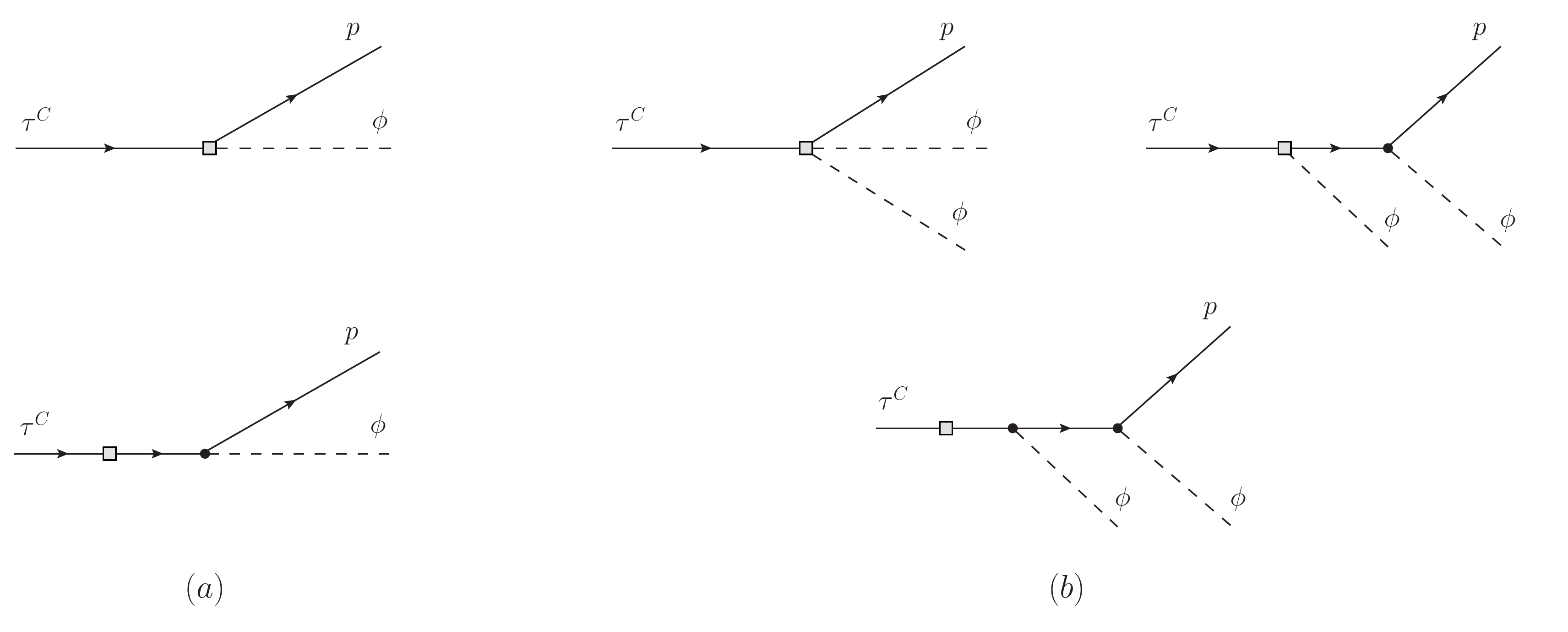} 
\caption{Generic diagrams that contribute to the amplitudes (a) $\tau^+\to p + \phi$ and (b) $\tau^+\to p+2\phi$.
The squares correspond to the lepton and baryon number violating interactions in
$\mathscr{L}_{B+L}^{had}$, while the circles are vertices from $\mathscr{L}_{B}$. The fermion in the 
intermediate lines is a proton. The amplitude for the process $\tau^+\to \Lambda \pi^+$ is also obtained 
from the diagrams in (a) with the external proton replaced by a $\Lambda$ baryon, $\phi=\pi^+$ and a
virtual $\Sigma^+$ in the fermion propagator.}
\label{fig:taudiagrams}
\end{centering}
\end{figure}
The generic diagrams contributing to the tau hadronic decay amplitude into one and two mesons,
arising from the vertices of $\mathscr{L}_{B+L}+\mathscr{L}_{B}$,
are shown in Figure~\ref{fig:taudiagrams}. 
In order to account for decays with an $\eta$-meson in the final state
we have to add a singlet contribution $\eta_1/\sqrt{3}\,\times\mathbb{I}$ to the pseudoscalar octet Eq.~(\ref{eq:u}).
The physical states $\eta$ and $\eta^\prime$ result from the mixing of the octet and singlet fields:
\begin{align}
\left(\begin{matrix}
 \eta_8\\
 \eta_1\\
\end{matrix}
\right)=\left(\begin{matrix}
              \cos\theta_P & \sin\theta_P\\
              -\sin\theta_P & \cos\theta_P\\
              \end{matrix}\right)
        \left(\begin{matrix}
              \eta\\
              \eta^\prime\\
              \end{matrix}\right) \,.
\end{align}
The large-$N_C$ limit of QCD yields a value for the $\eta-\eta^\prime$ mixing angle $\theta_P\simeq -20^\circ$~\cite{HerreraSiklody:1997kd},
which we use for the numerical results of Table~\ref{tab:rates}. Since phenomenological determinations of $\theta_p$ suggest values
ranging between $-10^\circ$ and  $-20^\circ$ we also provide results for the tau decay rates to $\eta$ mesons as a function
of $\theta_P$ in Table~\ref{tab:rateseta}.
\par
\begin{figure}[t!]
\begin{centering}
\includegraphics[width=1.\textwidth]{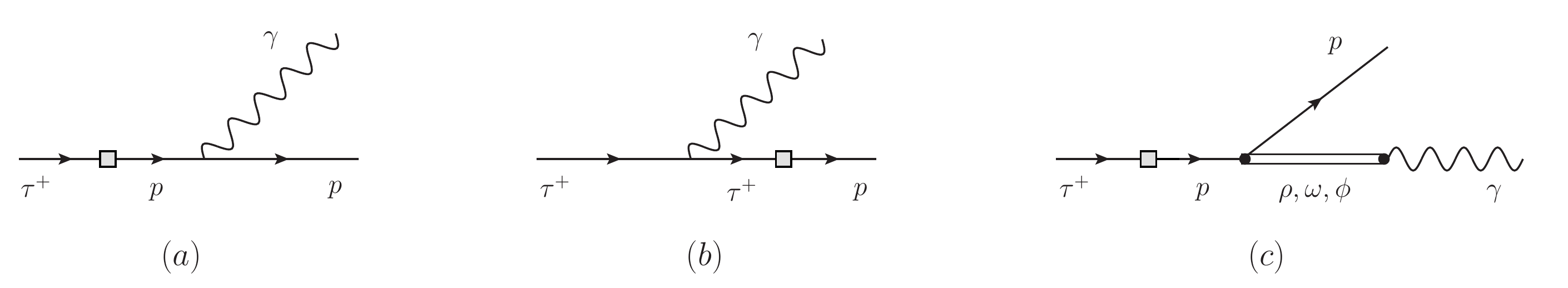} 
\caption{Diagrams that contribute to the lowest-order
amplitude for $\tau^+\to p \gamma$.
The squares correspond to the lepton and baryon number violating interactions in
$\mathscr{L}_{B+L}^{had}$. Proton and resonance interactions in diagram (c) are described by the
terms in $\mathscr{L}_{BV}$, Eq.~(\ref{eq:BV}), while the coupling of resonances to the photon is
contained in $\mathscr{L}_{2}(V)$, Eq.~(\ref{eq:L2V}).
The amplitude for the process 
$\tau^+\to p\mu^+\mu^-$ is also obtained from these diagrams 
by attaching a muon current to the (off-shell) photon.}
\label{fig:emdecay}
\end{centering}
\end{figure}

The electromagnetic decays $\tau^+ \to p\gamma,\,p\mu^+\mu^-$ proceed through the coupling of
the photon to the nucleon via an intermediate vector meson (see
Figure~\ref{fig:emdecay}c). This is because the amplitudes for diagrams where the photon 
couples directly to the fermion charge
through the covariant derivative in the kinetic term, Figures~\ref{fig:emdecay}a 
and~\ref{fig:emdecay}b, cancel each other.\footnote{The coupling of the photon to the charge of the
nucleon is readily obtained by taking $r_\mu=l_\mu=e Q A_\mu$ in Eq.~(\ref{eq:covder}), with
$Q=\frac{1}{3}diag\left(2,-1,-1\right)$ the quark charge matrix.}
The interactions between the baryons and the vector mesons 
can also be written in terms of a chirally-invariant Lagrangian. To lowest order
in the number of derivatives, it reads~\cite{Borasoy:1995ds,Kubis:2000zd}:
\begin{align}\label{eq:BV}
\mathscr{L}_{BV}=R_F\langle\overline{B}\sigma^{\mu\nu}\left[V_{\mu\nu},B\right]\rangle
+R_D\langle\overline{B}\sigma^{\mu\nu}\left\{V_{\mu\nu},B\right\}\rangle
+R_S\langle\overline{B}\sigma^{\mu\nu}B\rangle\langle V_{\mu\nu}\rangle
\,,
\end{align}
where 
\begin{align}
V_{\mu\nu}=\left(\begin{matrix}
                 \frac{\rho^0}{\sqrt{2}}+\frac{\omega_8}{\sqrt{6}}& \rho^+ & K^{*+}\\
                 \rho^- & -\frac{\rho^0}{\sqrt{2}}+\frac{\omega_8}{\sqrt{6}} & K^{*0}\\
                 K^{*-} & \overline{K}^{*0} & -\frac{2\omega_8}{\sqrt{6}}\\
                 \end{matrix}\right)_{\mu\nu}
\,,
\label{eq:Vmunu}
\end{align}
is the octet of spin-1 mesons, which  transforms in the same way as the baryon octet under chiral 
transformations, {\it i.e.} $V_{\mu\nu}\stackrel{G}{\to}h\left(g,\phi\right)V_{\mu\nu}\,h\left(g,\phi\right)^{-1}$.
Vector mesons in Eq.~(\ref{eq:Vmunu}) have been written in terms of antisymmetric tensor fields. The 
free Lagrangian in this formalism is given by~\cite{Ecker:1988te}:
\begin{align}
\mathscr{L}_{K}(V)=-\frac{1}{2}\langle\nabla^\lambda V_{\lambda\mu}\nabla_\nu V^{\nu\mu}
-\frac{M_V^2}{2}V_{\mu\nu}V^{\mu\nu}\rangle\,,
\end{align}
with $M_V$ the octet mass in the limit where the chiral symmetry is exact.
The physical $\omega$ and $\phi$ resonances are a superposition of an octet component, $\omega_8$, and a singlet one,
$\omega_1$, which can be added as a diagonal matrix $\omega_1/\sqrt{3}\,\mathbb{I}$ to the octet, Eq.~(\ref{eq:Vmunu}). We shall assume
ideal mixing; the fields in the Lagrangian are then expressed in terms of the physical fields as:
\begin{align}
\omega_{8,\mu\nu}=\frac{1}{\sqrt{3}}\,\omega_{\mu\nu}+\sqrt{\frac{2}{3}}\,\phi_{\mu\nu}\,, \quad 
\quad \omega_{1,\mu\nu}=\sqrt{\frac{2}{3}}\,\omega_{\mu\nu}-\frac{1}{\sqrt{3}}\,\phi_{\mu\nu} \,.
\end{align}
The actual couplings of the proton to the $\rho,\,\omega$ and $\phi$ mesons are proportional
to the combinations $(R_D+R_F)$, $(R_D+R_F+2R_S)$ and $(R_D-R_F+R_S)$, respectively. For the numerical
evaluation we have used the values obtained in Ref.~\cite{Mergell:1995bf}.
Finally, the lowest-order interactions of resonances with Goldstone boson fields as well as
external vector and axial-vector sources can be written as:
\begin{align}\label{eq:L2V}
\begin{aligned}
\mathscr{L}_{2}(V)&=\frac{F_V}{2\sqrt{2}}\langle V_{\mu\nu}f^{\mu\nu}_+\rangle
+\frac{iG_V}{\sqrt{2}}\langle V_{\mu\nu}u^\mu u^\nu\rangle
\,,
\\[3mm]
f_\pm^{\mu\nu}&=uF_L^{\mu\nu}u^\dagger \pm u^\dagger F_R^{\mu\nu}u
\,,
\end{aligned}
\end{align}
where $F_{R,L}$ are the field strength tensors of the left, $l_\mu$, and right, $r_\mu$, external sources and 
$F_V$ and $G_V$ are real couplings. The interaction between the resonances and the photon
is contained in the  operator with coefficient $F_V$ since
$f_+^{\mu\nu}=2eQF^{\mu\nu}+\dots$
with $F^{\mu\nu}$ the electromagnetic strength tensor and 
$Q=\frac{1}{3}diag\left(2,-1,-1\right)$ the quark charge matrix. We adopt 
the phenomenological value $F_V \simeq 154$~MeV~\cite{Ecker:1988te} for the numerics.
For the computation of the $\tau^+\to p \mu^+\mu^-$ decay rate, we need to introduce the
resonance widths to avoid the pole singularities in the phase-space integration over
the invariant mass of the lepton pair. This is done by using Breit-Wigner propagators for the 
virtual resonances in Figure~\ref{fig:emdecay}c, with a fixed width for the narrow $\omega$ and $\phi$
resonances, and the $q^2$-dependent width derived in~\cite{GomezDumm:2000fz} 
for the $\rho$ resonance.

Finally, the results obtained for the tau decay rates have the form:  
\begin{align}
\Gamma = \frac{1}{\Lambda^4} \, \Big[ & \,
  a_1 \, \big( |C^\prime_{RL}|^2 + |C^\prime_{LR}|^2 \big)
+ a_2 \, {\rm Re}\,\left\{ C^\prime_{RL}\,C^{\prime\,*}_{LR} \right\} 
+ a_3 \, \big( | C^\prime_{RR}|^2 + | C^\prime_{LL}|^2 \big)
+ a_4 \, {\rm Re}\,\left\{ C^\prime_{RR}\, C^{\prime\,*}_{LL} \right\}
\nonumber\\[2mm]
&
+ a_5 \, {\rm Re}\,\big( C^\prime_{RL}\,C^{\prime\,*}_{LL} + C^\prime_{LR}\,C^{\prime\,*}_{RR} \big)
+ a_6 \, {\rm Re}\,\big( C^\prime_{RL}\,C^{\prime\,*}_{RR} + C^{\prime}_{LR}\,C^{\prime\,*}_{LL} \big) 
\, \Big] 
\,,
\label{eq:rates}
\end{align}
where we have included the strong coefficients $\alpha,\,\beta$ in the definition of the primed coefficients:
\begin{align}
\begin{aligned}
C_X^\prime \equiv \alpha \, C_X\,, \quad&\quad X=RL,LR \,,
\\
C_X^\prime \equiv \beta \, C_X\,, \quad&\quad X=LL,RR  \,,
\end{aligned}
\end{align}
and equivalently for the $\tilde{C}_X$, which are only relevant for the $\Delta S=1$ decay 
$\tau^+ \to \Lambda \pi^+$.
The numerical values obtained for the coefficients $a_i$ in Eq.~(\ref{eq:rates}) have been collected 
in Tables~\ref{tab:rates} and~\ref{tab:rateseta}. The analytic expressions for the tau decays
are lengthy and not very illuminating; only for the case of two particles in the final state, the leading order term in
the expansion in the mass of the pseudoscalar mesons gives a concise formula. This is the case 
for the process with largest $a_i$ coefficients, $\tau\to p\,\pi^0$:
\begin{align}\nonumber
\Gamma(\tau^+\to p\,\pi^0) = &\frac{(m_\tau^2-m_p^2)}{128\pi f_\pi^2\, m_\tau \, \Lambda^4}\,
\bigg\{
 \frac{4\,m_p}{m_\tau} \, \left[ 1- (D+F)^2 \right] \,
\mbox{Re} \left\{ (C^\prime_{RL}+C^\prime_{LL})\,  (C^{\prime}_{LR}+C^{\prime}_{RR})^* \right\}
\\[2mm]\nonumber
&+ \, \Big( \big| C^\prime_{RL}+C^\prime_{LL} \big|^2  +  \big| C^\prime_{LR}+C^\prime_{RR} \big|^2 \Big)
 \left[ (1+D+F)^2 + \frac{m_p^2}{m_\tau^2}\, (1-D-F)^2  \right]\\[2mm]
&+{\cal O}\Big(\frac{m_{\pi}^2}{m_\tau^2} \Big)\bigg\}
\,.
\label{eq:tauppi0decay}
\end{align}
For the similar decay $\tau\to p\,\eta$, however, the analytic result becomes
more cumbersome due to the $\eta^0-\eta^8$ mixing and shall not be given here. 
On the other hand,
the decay rate formula for the process $\tau\to \Lambda\,\pi^+$, which has the more stringent experimental 
bound among the $\Delta B=\Delta L=1$ tau decays, also acquires a simple
form if we neglect the mass difference
between the $\Lambda$ and the $\Sigma$ baryons, namely:
\begin{align}
\begin{aligned}
\Gamma(\tau^+\to \Lambda\,\pi^+) &= \frac{(m_\tau^2-m_\Lambda^2)}{96\pi f_\pi^2\, m_\tau \, \Lambda^4}\,
\bigg\{
-\frac{4\,m_p}{m_\tau} \, D^2 \,
\mbox{Re} \left\{ (C^\prime_{RL}+C^\prime_{LL})\,  (C^{\prime}_{LR}+C^{\prime}_{RR})^* \right\}
\\[2mm]
+& \,  \Big( 1 -\frac{m_{\pi}^2}{m_\tau^2}  \Big)  \,2D\,
\mbox{Re} \left\{ C^\prime_{LR} \, (C^\prime_{RL}+C^\prime_{LL})^* + C^\prime_{RL} \, (C^{\prime}_{LR}+C^{\prime}_{RR})^* \right\}
\\[2mm]
+& \,\Big( 1+ \frac{m_{\pi}^2}{m_\tau^2}  \Big) \, 
\left[ D^2\,\left( \,\big| C^\prime_{RL}+C^\prime_{LL} \big|^2  +  \big| C^\prime_{LR}+C^\prime_{RR} \big|^2 \, \right)
+  \big| C^\prime_{LR} \big|^ 2+  \big| C^\prime_{RL} \big|^2 \right]
\\[2mm]
+&  \frac{4m_{\pi}}{m_\tau}  \, \mbox{Re} \left\{ C^\prime_{LR} \, C^{\prime\,*}_{RL} \right\}+
\, {\cal O}\Big(\frac{m_{\pi}^2}{m_\tau^2} \Big) + {\cal O}\Big(m_\Lambda - m_\Sigma \Big)  \bigg\}
\,.
\end{aligned}
\end{align}
In Tables~\ref{tab:rates} and~\ref{tab:rateseta}, the values used for all particle masses, 
as well as for the $\phi$- and $\omega$-resonance widths, correspond to those listed in the PDG~\cite{Beringer:1900zz}.

\renewcommand{\arraystretch}{2.}
\setlength{\tabcolsep}{7pt}
\begin{table}[t]
\centering
\begin{tabular}{| c | c c c c c c |}
 \hline
& $a_1$ & $a_2$ & $a_3$ & $a_4$ & $a_5$ & $a_6$
  \\
 \hline\hline 
 $\Gamma(\tau^+  \to p\,\pi^0)$ & 1.87 & -0.419  & 1.87 & -0.419 & 3.74 & -0.419
    \\
 \hline
$\Gamma(\tau^+ \to p\,\eta)$ & 0.130 & -0.181 & 1.38 & 1.56 & 0.654 & -0.0705
  \\
 \hline
 $\Gamma(\tau^+ \to p\, \pi^0\pi^0)$ & 0.124 & 0.0481 & 0.124 & 0.0481 & 0.247 & 0.0481
 \\
  \hline  
$\;\Gamma(\tau^+ \to p\, \pi^0\eta) \times 10^{2}\; $ & 0.0874  & 0.0322 & 1.87 & -0.262 & 0.689 & -0.000549
 \\
  \hline
$\;\Gamma( \tau^+ \to p \,\gamma) \times 10^{3}\;$ & 3.60 & -5.95 & 3.60 & -5.95 & 7.21 & -5.95
  \\
 \hline
$\;\Gamma(\tau^+ \to p\, \mu^+\mu^-)\times 10^{5}\;$ & 1.26 & -1.49 & 1.26 & -1.49 & 2.53 & -1.49
  \\
 \hline
$\;\Gamma(\tau^+ \to \Lambda \, \pi^+)\;$ & 1.41 & 0.173 &  0.440 & -0.811 & 1.29 & -0.832
  \\
 \hline
\end{tabular}
\caption{Coefficients in Eq.~(\ref{eq:rates}) in units of GeV${}^{-1}$ for different rates. In the case
of $p\,\eta$ and $p\,\pi^0\eta$ final states we have used $\theta_P \simeq -20^\circ$ for the $\eta-\eta^\prime$ 
mixing angle.}
\label{tab:rates}

\end{table}

\renewcommand{\arraystretch}{2.}
\setlength{\tabcolsep}{7pt}
\begin{table}[H]
\centering
\begin{tabular}{| c | c | c |}
 \hline
& $\Gamma(\tau^+ \to p\,\eta)$ & $\Gamma(\tau^+ \to p\, \pi^0\eta) \times 10^{2}$
  \\
 \hline\hline 
$a_1$ 
& 
$ 1.90 \,s_\theta^2 + 0.0741 \, c_\theta^2  + 0.490 \,s_\theta \,c_\theta $
& 
$ 2.39 \,s_\theta^2 + 0.101 \, c_\theta^2  + 0.877 \,s_\theta \,c_\theta $ 
  \\
 \hline
$a_2$ 
& 
$ 1.35 \,s_\theta^2 + 0.129 \, c_\theta^2  + 1.41 \,s_\theta \,c_\theta $ 
&                                              
$ -0.244 \,s_\theta^2 - 0.0133 \,c_\theta^2 - 0.226 \,s_\theta \,c_\theta $ 
  \\
 \hline
$a_3$ 
& 
$
 0.370 \,s_\theta^2 + 1.19 \, c_\theta^2   - 0.886 \,s_\theta \,c_\theta $
& 
$ 0.242 \,s_\theta^2 + 1.71 \,c_\theta^2  - 1.03 \,s_\theta \,c_\theta $
  \\
 \hline
$a_4$ 
& 
$ -0.536 \,s_\theta^2 + 1.81 \, c_\theta^2 - 0.0744 \,s_\theta \,c_\theta $
& 
$ 0.0965 \,s_\theta^2 - 0.286\,c_\theta^2 + 0.0628 \,s_\theta \,c_\theta $
  \\
 \hline
$a_5$ 
& $ 1.40 \,s_\theta^2 - 0.487 \, c_\theta^2  - 2.87 \,s_\theta \,c_\theta $
& 
$ 1.33 \,s_\theta^2 - 0.789\,c_\theta^2  - 3.83 \,s_\theta \,c_\theta $
  \\
 \hline
$a_6$ 
& 
$ -0.371 \,s_\theta^2 - 0.590 \, c_\theta^2  - 1.54 \,s_\theta \,c_\theta $
& 
$ 0.0435  \,s_\theta^2 + 0.0846 \,c_\theta^2  + 0.250 \,s_\theta \,c_\theta $
  \\
 \hline
\end{tabular}
\caption{Coefficients in Eq.~(\ref{eq:rates}) in units of GeV${}^{-1}$ for the 
$\tau^+ \to p\,\eta$ and $\tau^+ \to p\,\pi^0\eta$ rates, as 
a function of the $\eta-\eta^\prime$ 
mixing angle ($s_\theta\equiv \sin \theta_P$ and $c_\theta\equiv \cos \theta_P$).}
\label{tab:rateseta}

\end{table}

\end{document}